\documentclass[10pt,journal,compsoc]{IEEEtran}

\usepackage[T1]{fontenc}
\usepackage[utf8]{inputenc}
\usepackage[american]{babel}
\usepackage{csquotes}
\usepackage{microtype}
\usepackage[htt]{hyphenat}
\usepackage{paralist}
\setdefaultenum{(1)}{(a)}{(i)}{A.}
\usepackage{enumitem}
\usepackage{url}
\usepackage{flushend}
\usepackage{etoolbox}
\usepackage{fix-cm}
\usepackage{textpos}
\usepackage{mfirstuc}
\usepackage{titlecaps}
\usepackage[datesep=.,style=ddmmyyyy]{datetime2}

\usepackage{algorithm}
\usepackage{algpseudocode}

\usepackage{amsmath}
\usepackage{amssymb}
\usepackage{bm}
\usepackage{relsize}
\usepackage{siunitx}
\usepackage[super]{nth}

\usepackage{booktabs}
\usepackage{multirow}
\usepackage{colortbl}
\usepackage{makecell}
\usepackage{rotating}
\usepackage{threeparttable}

\usepackage{float}
\usepackage{graphicx}
\usepackage[table,dvipsnames]{xcolor}

\usepackage{listingsutf8}

\usepackage{xparse}
\usepackage{xspace}

\usepackage[xindy,acronym]{glossaries}
\glsdisablehyper

\usepackage[numbers,sort&compress]{natbib}

\usepackage[hidelinks,bookmarks=false]{hyperref}
\usepackage[all]{hypcap}

\usepackage[capitalise]{cleveref}

\newcommand{\vda}{$\hat{A}_{12}$}
\newcommand{\code}[1]{\texttt{#1}}
\newcommand{\tool}[1]{\textit{#1}}
\newcommand{\oasType}[1]{\textit{#1}}

\newcommand{\version}[1]{\texttt{#1}}

\newcommand{\evomaster}{\tool{EvoMaster}}
\newcommand{\evomasterbb}{\tool{\evomaster-BB}}
\newcommand{\evomasterwb}{\tool{\evomaster-WB}}
\newcommand{\evomasterwbmio}{\tool{\evomasterwb-MIO}}
\newcommand{\evomasterwbmosa}{\tool{\evomasterwb-MOSA}}
\newcommand{\evomasterwbwts}{\tool{\evomasterwb-WTS}}
\newcommand{\evoguri}{\tool{EvoGURI}}
\newcommand{\evoguriRandom}{\tool{\evoguri}}

\newcommand{\evosuite}{\tool{EvoSuite}}
\newcommand{\guri}{\tool{GURI}}
\newcommand{\numVersions}{\num{10}}
\newcommand{\oasStrict}{\oasType{strict}}
\newcommand{\oasDefault}{\oasType{default}}

\newcommand{\cov}[1]{#1}
\newcommand{\covLine}{\cov{Line}}
\newcommand{\covBranch}{\cov{Branch}}
\newcommand{\covMethod}{\cov{Method}}
\newcommand{\numMutations}{\num{2279}}
\newcommand{\numMutationOperators}{\num{8}}
\newcommand{\numCancerMessageVariables}{\num{51}}

\newcommand{\guriRule}[1]{#1}
\newcommand{\validationRule}[1]{\guriRule{V#1}}

\newcommand{\ruleResult}[1]{#1}
\newcommand{\rulePass}{\ruleResult{Pass}}
\newcommand{\ruleFail}{\ruleResult{Fail}}
\newcommand{\ruleWarning}{\ruleResult{Warning}}
\newcommand{\ruleNotApplied}{\ruleResult{Not Applied}}

\newcommand{\mut}[1]{\textit{#1}}
\newcommand{\mutAD}{\mut{AD}}
\newcommand{\mutNI}{\mut{NI}}
\newcommand{\mutRI}{\mut{RI}}
\newcommand{\mutSComp}{\mut{SComp}}
\newcommand{\mutSConn}{\mut{SConn}}
\newcommand{\mutSSE}{\mut{SSE}}
\newcommand{\mutSSR}{\mut{SSR}}
\newcommand{\mutSSI}{\mut{SSI}}

\newcommand{\rqcodecov}{RQ1}
\newcommand{\rqfaults}{RQ2}
\newcommand{\rqexecutedrules}{RQ3}

\newcommand{\rqrulefrequency}{RQ4}
\newcommand{\rqmut}{RQ5}

\newcommand{\nrcols}[2]{\multicolumn{#1}{l}{#2}}

\definecolor{greylight}{RGB}{240,240,240}
\definecolor{greymedium}{RGB}{189,189,189}
\definecolor{greydark}{RGB}{99,99,99}

\usepackage{tcolorbox}
\tcbset{colback=greylight, colframe=greydark, arc=1pt, boxrule=0.75pt}
\newcommand{\summarybox}[2]{
	\begin{tcolorbox}
		\small{
			\textbf{#1:} #2
		}
	\end{tcolorbox}
}
 
\newacronym{api}{API}{application programming interface}
\newacronym{caress}{CaReSS}{Cancer Registration Support System}
\newacronym{cpu}{CPU}{central processing unit}
\newacronym{crn}{CRN}{Cancer Registry of Norway}
\newacronym{db}{DB}{database}
\newacronym{dto}{DTO}{data transfer object}
\newacronym{ea}{EA}{evolutionary algorithm}
\newacronym{ehr}{EHR}{electronic health record}
\newacronym{ex3}{eX\textsuperscript{3}}{Experimental Infrastructure for Exploration of Exascale Computing}
\newacronym{ga}{GA}{genetic algorithm}
\newacronym{gan}{GAN}{generative adversarial network}
\newacronym{gui}{GUI}{graphical user interface}
\newacronym{hpc}{HPC}{high-performance computing}
\newacronym{html}{HTML}{HyperText Markup Language}
\newacronym{http}{HTTP}{Hypertext Transfer Protocol}
\newacronym{io}{I/O}{input-output}
\newacronym{jar}{JAR}{Java Archive}
\newacronym{jdk}{JDK}{Java Development Kit}
\newacronym{json}{JSON}{JavaScript Object Notation}
\newacronym{jvm}{JVM}{Java virtual machine}
\newacronym{llm}{LLM}{large language model}
\newacronym{loc}{LOC}{lines of code}
\newacronym{maoea}{MaOEA}{many-objective evolutionary algorithm}
\newacronym{mio}{MIO}{Many Independent Objective}
\newacronym{ml}{ML}{machine learning}
\newacronym{moea}{MOEA}{multi-objective evolutionary algorithm}
\newacronym{mosa}{MOSA}{Many-Objective Sorting Algorithm}
\newacronym{ms}{MS}{mutation score}
\newacronym{oas}{OAS}{OpenAPI Specification}
\newacronym{ocl}{OCL}{Object Constraint Language}
\newacronym{pp}{pp}{percentage points}
\newacronym{rcn}{RCN}{The Research Council of Norway}
\newacronym{rest}{REST}{representational state transfer}
\newacronym{rmit}{RMIT}{randomized multiple interleaved trials}
\newacronym{rpc}{RPC}{remote procedure call}
\newacronym{rq}{RQ}{research question}
\newacronym{sql}{SQL}{structured query language}
\newacronym{sut}{SUT}{subject under test}
\newacronym{uml}{UML}{Unified Modeling Language}
\newacronym{vae}{VAE}{variational autoencoder}
\newacronym{wts}{WTS}{Whole Test Suite}
 
\begin{document}

 \title{Testing Medical Rules Web Services in Practice}

\author{
  Christoph~Laaber, Shaukat~Ali, Thomas~Schwitalla, and
  Jan~F.~Nygård \IEEEcompsocitemizethanks{\IEEEcompsocthanksitem C. Laaber is with the Simula Research Laboratory, Oslo, Norway\protect\\
    E-mail: laaber@simula.no
    \IEEEcompsocthanksitem S. Ali is with the Simula Research Laboratory and the Oslo Metropolitan University, Oslo, Norway\protect\\
    E-mail: shaukat@simula.no
    \IEEEcompsocthanksitem T. Schwitalla is with the Cancer Registry of Norway, Norwegian Institute of Public Health, Oslo, Norway\protect\\
    E-mail: thsc@kreftregisteret.no
    \IEEEcompsocthanksitem J. F. Nygård is with the Cancer Registry of Norway, Norwegian Institute of Public Health, Oslo, Norway and UiT The Arctic University of Norway, Tromsø, Norway\protect\\
    E-mail: jfn@kreftregisteret.no
  }}
 
\IEEEtitleabstractindextext{
\glsresetall{}

\begin{abstract}

The \gls{crn} collects and processes cancer-related data for patients in Norway.
For this, it employs a sociotechnical software system that evolves with changing requirements and medical standards.
The current practice is to manually test \gls{crn}'s system to prevent faults and ensure its dependability.
This paper focuses on automatically testing \guri{}, the \gls{crn}'s medical rule engine, using a system-level testing tool, \evomaster{}, in both its black-box and white-box modes, and a novel \gls{crn}-specific \evomaster-based tool, \evoguri{}.
We empirically evaluate the tools' effectiveness regarding code coverage, errors found, domain-specific rule coverage, and ability to identify artificial faults ten versions of \guri{}.
Our results show that all the tools achieve similar code coverage and identified a similar number of errors.
For rule coverage, \evoguri{} and \evomaster{}'s black-box mode produce test suites that cover the highest number of rules with \rulePass{}, \ruleFail{}, and \ruleWarning{} results.
The test suites of \evoguri{} and two \evomaster{} white-box tools identify the most faults in a mutation testing experiment.
Based on our findings, we recommend using \evoguri{} in \gls{crn}'s current practice.
Finally, we present key takeaways and outline open research questions for the research community.

\end{abstract}

}

\maketitle

\glsresetall{}

\section{Introduction}

Cancer is a leading cause of death worldwide, with nearly \num{10} million deaths in 2022~\citep{bray:24}.
Consequently, most countries systematically collect data about cancer patients in specialized registries to improve patient care by supporting decision-making and conducting research.
These registries maintain specialized software systems to collect, curate, and analyze cancer data.
However, engineering such software systems poses many challenges, such as
\begin{inparaenum}
    \item collecting data for patients throughout their lives from diverse sources, e.g., hospitals, laboratories, and other registries;
    \item dealing with continuous evolution, e.g., due to software updates, new requirements, updated regulations, and new medical research;
    and
    \item increased incorporation of machine learning algorithms to decision support and production of statistics for relevant stakeholders, including patients and policymakers.
\end{inparaenum}

Our context is the \gls{crn}, which has developed a \gls{caress} to support collecting, processing, and managing cancer-related data from various medical entities such as Norwegian hospitals and laboratories.
Based on the processing, the \gls{crn} generates statistics that are consumed by external entities, including policymakers, hospitals, and patients.
It further provides data for researchers to conduct research.
Naturally, the quality of statistics and data depends on how correct and reliable \gls{caress} is.
To this end, testing is one method to ensure the dependability of \gls{caress}.

This paper reports on a case study in the real-world context of the \gls{crn}, focusing on testing one of the key components of \gls{caress}, called \guri{} --- a rule engine responsible for checking medical rules for various purposes such as data validation and aggregation.
Currently, \guri{} is primarily tested manually, as also reported by \citet{haas:21} that manual testing is a common practice in industry.
This paper investigates test generation tools to enable automated testing to alleviate manual testing of \guri{}.
For this, our study takes a popular open-source, AI-based system-level test generation tool called \evomaster{}~\citep{arcuri:19}, which has been shown to be superior to nine other \gls{rest} \gls{api} testing tools~\citep{kim:22a}, addresses domain-specific challenges by devising a novel \gls{crn}-specific \evomaster{} extension called \evoguri{}, and performs automated testing of \guri{} with these to assess the testing effectiveness of \guri{} from various perspectives.

In our previous conference paper~\citep{laaber:23a}, we applied \evomaster{}'s four off-the-shelf tools, i.e., black-box and white-box tools with three \glspl{ea}, to test \guri{}'s \numVersions{} versions, relying on \guri{}'s \oasDefault{} \gls{oas}.
We assessed \evomaster{}'s effectiveness with three metrics: source code coverage, errors found, and domain-specific rule coverage.
We found that \evomaster{} struggles to cover specific rules, limiting its applicability for the \gls{crn}.
To address this limitation, this paper extends the previous paper by
\begin{inparaenum}
    \item extending \evomaster{} with domain-specific rule objectives and targets into the \evoguri{} tool;
    \item assessing the tools with two \gls{oas} types, i.e., the \oasDefault{} and \oasStrict{} version with added variable constraints;
    and
    \item strengthening the empirical evaluation through a mutation testing experiment comprising \numMutations{} rule mutations from \numMutationOperators{} rule mutation operators.
\end{inparaenum}

The results of our experiments show that all five tools' effectiveness is similar in terms of code coverage and errors found across all \guri{} versions.
However, we observe that \evoguri{} and \evomaster{}'s black-box tool generate test suites that execute more rules leading to \rulePass{}, \ruleFail{}, and \ruleWarning{} results.
During the test generation phase, the \evomaster{} white-box tool with the \gls{mio} algorithm executes rules with \rulePass{}, \ruleFail{}, and \ruleWarning{} results with the highest frequency compared to all executed rules, followed by \evoguri{} and the other \evomaster{} white-box tools.
We further compare the rule executions observed by \guri{} in production, and the results show that no tool can \rulePass{} rules as often as observed in production, but they all are good at triggering \ruleFail{} and \ruleWarning{} results, which are the exceptional and corner cases one would predominantly like to test.
In terms of mutation testing, \evoguri{} and the \evomaster{} white-box tools with the \gls{mosa} and \gls{wts} algorithms achieve \glspl{ms} between \num{0.96} and \num{1}, showing that the generated test suites are highly effective in identifying artificial seeded faults.
Our results do not suggest that rule evolution plays a critical role, as the tools' effectiveness remains stable across the \numVersions{} rule versions.
Finally, we notice that relying on a \oasStrict{} \gls{oas} with precise variable constraints improves the tools' effectiveness across all the tools and studied metrics.

Based on our results, we recommend using \evoguri{} as the automated testing solution for \guri{} and \gls{caress}.
In particular, this paper compared to our previous conference paper~\citep{laaber:23a} shows that our domain-aware \evoguri{} combined with a \oasStrict{} \gls{oas} leads to vastly improved effectiveness, covering \num{0.4981515712} \gls{pp} more \rulePass{} and \num{0,6049907579} \gls{pp} more \ruleFail{} results.
Finally, we provide detailed discussions and lessons from our industrial case study regarding its generalization to other contexts and point out key research areas that deserve attention from the software engineering community.

\section{Application Context}
\label{sec:background:context}
The \gls{crn} regularly collects cancer patients' data (e.g., diagnostic, treatment), based on which cancer research and statistics can be conducted.
To ensure the quality of the collected data, the \gls{crn}'s \gls{caress} has introduced several preventive efforts to discover and amend inaccurate or missing data.
Patient data is submitted to the \gls{crn} as \textit{cancer messages}, which contain the patient data as \numCancerMessageVariables{} medical variables.
Cancer messages are coded by medical personell in a stringent process and aggregated into \textit{cancer cases}.
Cancer messages and cases represent a timeline of a patient's diagnoses, treatments, and follow-ups.
The coding process relies on standard classification systems and depends on hundreds of \textit{medical rules} for validating cancer messages and aggregating cancer cases.
Consequently, these rules are of two types: \textit{validation rules} and \textit{aggregation rules}, which are defined by medical experts and implemented in \guri{} for automated validation and aggregation of cancer messages and cases.
These rules constantly evolve, e.g., due to updated medical knowledge and procedures.

Below is a validation rule, which states that for all the cancer messages of type \textit{H}, if the \textit{surgery} value is equal to \num{96}, then the \textit{basis} value must be greater than \num{32}. 
\begin{equation*}
	\forall \; messageType = H \implies (surgery = 96 \implies basis > 32)
\end{equation*}

The aggregation rule below determines the state of \emph{Morphologically verified}, based on a given \emph{Basis} value of cancer messages:
\vspace{-1em}
\begin{algorithm}[ht]
    \small
    \begin{algorithmic}[0]
   \If{Basis $\in$ ['22', '32', '33', '34', '35', '36', '37', '38', '39', '57', '60', '70', '72', '74', '75', '76', '79']}
   \State Morphologically verified = 'Yes'
    \ElsIf{Basis $\in$ ['00', '10', '20', '23', '29', '30', '31', '40', '45', '46', '47', '90', '98']}
    \State Morphologically verified = 'No'
    \Else 
    \State Morphologically verified = null
    \EndIf
\end{algorithmic}
\end{algorithm}

Both rule types can yield four result types for a given cancer message or case:
\begin{inparadesc}
    \item [\rulePass{}] results from a successful rule execution, i.e., the cancer message with all its data is valid, or the cancer case was successfully aggregated with a previous cancer case and a number of cancer messages; 
    \item [\ruleFail{}] results from an unsuccessful rule execution, i.e., the cancer message is invalid, or the cancer case fails to be aggregated;
    \item [\ruleWarning{}] is the result where the rule execution is successful; however, the data appears to be dubious (e.g., a patient's age is \num{120} years, which is theoretically valid but highly unlikely);
    and
    \item [\ruleNotApplied{}] is a rule that is partially executed against an input, which only applies to validation rules of the form $apply \implies rule$, where $apply$ is the condition on which $rule$ is applied (in the example above: a cancer message with $messageType != H$, the rule would result in \ruleNotApplied{}).
\end{inparadesc}

\guri{} is a component of \gls{caress} that stores the rules in a database, exposes a \gls{rest} \gls{api} (internally at the \gls{crn}), receives cancer messages and cases for validation and aggregation, and returns the result.
Upon receiving a cancer message or case, \guri{} applies all rules and returns the result for each rule.
New rules and updates to the rules are done through a \gls{gui}, which is available to medical personnel.

\section{Experimental Study}
\label{sec:study}

We perform laboratory experiments~\citep{stol:18} of automated test generation techniques for \gls{rest} \glspl{api} at the \gls{crn} to understand their effectiveness of covering source code; revealing errors; executing domain-specific elements, i.e., medical rules; and achieving high rule mutation scores.
Our experiment comprises a real-world study subject, i.e., \gls{crn}['s] medical rule engine \guri{}, and five \gls{rest} \gls{api} test generation tools.

\subsection{Research Questions}
\label{sec:study:rqs}

Our study investigates the following five \glspl{rq} to assess the effectiveness of the test generation tools:

\begin{description}[topsep=0.8em,labelindent=1em,labelwidth=3em,itemsep=0.2em]
    \item[\rqcodecov{} (code coverage)] How much code coverage do the tools achieve?
    \item[\rqfaults{} (code errors)] How many code-related errors do the tools trigger?
    \item[\rqexecutedrules{} (executed rules and results)] How many rules can the tools execute and which results do they yield?
    \item[\rqrulefrequency{} (rule and result frequency)] How often do the executed rules yield each result type during test generation, and how does it compare to production \guri{}?
    \item[\rqmut{} (mutation testing)] How effective are the generated tests in finding rule errors?
\end{description}

\rqcodecov{} and \rqfaults{} are \enquote{traditionally} investigated research questions for evaluating test generation tools in terms of effectiveness, including \gls{rest} \gls{api} test generation~\citep{kim:22a}. 
\rqexecutedrules{} and \rqrulefrequency{} are domain-specific \glspl{rq} that evaluate the test generation tools' effectiveness in testing \guri{}'s main functionality, i.e., validating and aggregating cancer messages and cancer cases.
\rqexecutedrules{} studies the test generation tools' capability to execute the medical rules and yield all possible result types per rule, with the ultimate goal of testing all rule-result pairs.
The goal of \rqrulefrequency{} is to investigate the frequency of each rule's result types during test generation and how these results compare to the results from the production system.
\rqmut{} applies the traditional concept of mutation testing to medical rules (instead of source code) to evaluate the generated test suites' ability in finding (artificial) rule faults.

\subsection{Study Subject --- \guri{}}
\label{sec:study:subject}

\guri{} is implemented as a Java web application with Spring Boot, which exposes \gls{rest} \gls{api} endpoints to \gls{crn}['s] internal systems and provides a web interface for medical coders.
Although \guri{} has \num{32} \gls{rest} endpoints, we only focus on \num{2} in this experiment, as these are the ones handling the rules, i.e., one for validating cancer messages with validation rules and one for aggregating cancer messages into cancer cases with aggregation rules.
\guri{}'s most recent version consists of \num{70} validation rules and \num{43} aggregation rules.
Since \guri{} was introduced, its source code has hardly changed; however, its rules have been subject to evolution due to updated medical knowledge, such as rule additions, deletions, and modifications.
Consequently, based on these changes, we form \numVersions{} rule sets as \numVersions{} versions in this experiment.
Specifically, out of \num{28} unique points in time when the rules were changed in \guri{}, we select \num{10} dates where the changes are most severe (the most rule additions and deletions occurred).
\Cref{tab:rule-versions} depicts this rule evolution.

\begin{table}
    \centering
    \caption{Rule Evolution}
    \label{tab:rule-versions}
    \begin{tabular}{llrr}
\toprule
Version & Date & \nrcols{2}{Rules} \\
\cmidrule(l{1pt}r{1pt}){3-4}
 &  & Validation & Aggregation \\
\midrule
\version{v1} & \DTMdate{2017-12-12} & \num{30} & \num{32} \\
\version{v2} & \DTMdate{2018-05-30} & \num{31} & \num{33} \\
\version{v3} & \DTMdate{2019-02-06} & \num{48} & \num{35} \\
\version{v4} & \DTMdate{2019-08-27} & \num{49} & \num{35} \\
\version{v5} & \DTMdate{2019-11-11} & \num{53} & \num{37} \\
\version{v6} & \DTMdate{2020-09-25} & \num{56} & \num{37} \\
\version{v7} & \DTMdate{2020-11-24} & \num{66} & \num{38} \\
\version{v8} & \DTMdate{2021-04-20} & \num{69} & \num{43} \\
\version{v9} & \DTMdate{2022-01-13} & \num{69} & \num{43} \\
\version{v10} & \DTMdate{2022-01-21} & \num{70} & \num{43} \\
\bottomrule
\end{tabular}
 \end{table}

\subsection{Test Generation Tools}
\label{sec:study:tools}

The automated \gls{rest} \gls{api} test generation tools form the independent variable of our experimental study.
We select \evomaster{}, a search-based software testing framework to generate system-level test suites for web \glspl{api} exposing \gls{http} endpoints among others, in version \version{v2.0.0}\footnote{\url{https://github.com/WebFuzzing/EvoMaster/releases/tag/v2.0.0}} with multiple parameterizations as the tools~\citep{arcuri:19}.
\evomaster{} was recently shown to be the most effective tool, in terms of source code coverage and triggered errors, among ten different tools~\citep{kim:22a}.

In terms of testing approach, our experiments use both black-box and white-box testing.
\evomasterbb{} relies on random testing, \gls{rest} request sampling based on an \gls{oas}, and an archive to store the generated tests that cover new targets~\citep{arcuri:21}.
The targets are the different \gls{http} endpoints and their status codes, e.g., successful (4xx status code) and faulty (5xx status code) request-response pairs.
On the other hand, \evomasterwb{} uses an \gls{ea} to generate tests relying on random sampling of \gls{rest} requests, coverage feedback, and mutation~\citep{arcuri:19,zhang:22a,marculescu:22}.
The \gls{ea} relies on the same targets as \evomasterbb{} as search objectives and additionally incorporates code coverage criteria such as line, branch, and method coverage.
\evomasterwb{} supports three \glspl{ea}:
\begin{inparaenum}
    \item \gls{mio}~\citep{arcuri:18b}, which is a \gls{maoea} focusing on scalability in the presence of many testing targets, that was specifically designed for \gls{rest} \gls{api} test generation and is \evomaster{}'s default;
    \item \gls{mosa}~\citep{panichella:18}, which is the first \gls{maoea} that was designed for unit test generation with \evosuite{}~\citep{fraser:11};
    and
    \item \gls{wts}~\citep{fraser:13}, which is a single objective \gls{ga} that was designed for unit test generation and is the original \gls{ea} of \evosuite{}.
\end{inparaenum}

In addition, influenced by our previous paper's findings~\citep{laaber:23a}, we extend \evomasterbb{} to include \gls{crn}-context-specific targets (see \cref{sec:background:context}).
We call this extension \evoguri{}, which leverages the returned result types for each rule as a covered target, in addition to \evomasterbb{}'s \gls{rest} endpoint and status code targets.
For each test case (i.e., \gls{http} request), \evoguri{} parses the response body into rule and result pairs, e.g., \validationRule{01} (validation rule number \num{1}) had the result \rulePass{}.
Note that all the rules are executed against the cancer message or case of each test case.
Each test case covers $n$ targets, one for each rule, for a specific \gls{rest} endpoint and status code.
\evoguri{} saves a test case in an external archive, if it covers new targets, i.e., it executes a rule that yields a new result.
Through this, \evoguri{} keeps test cases with uncovered rule-result-pairs with the goal of generating a comprehensive test suite that improves rule fault discovery compared to the standard \evomaster{} test suites.

Our experiment investigates all four \evomaster{} tool parameterizations and \evoguri{}, which we simply call \enquote{tool} in the remainder of the paper:
\begin{inparaenum}
    \item \textbf{\evomasterbb{}:} \evomaster{} black-box;
    \item \textbf{\evomasterwbmio{}:} \evomaster{} white-box with \gls{mio};
    \item \textbf{\evomasterwbmosa{}:} \evomaster{} white-box with \gls{mosa};
    \item \textbf{\evomasterwbwts{}:} \evomaster{} white-box with \gls{wts};
    and
    \item \textbf{\evoguri{}:} \evomasterbb{} with rule-result targets.
\end{inparaenum}
Beyond the testing approach and the \gls{ea}, our experiment uses the default parameters of \evomaster{}, similar to \citet{kim:22a}.

\subsection{\Gls{oas} Types}
\label{sec:study:oas}

In our previous paper~\citep{laaber:23a}, we relied on the \gls{oas} that is generated by \guri{} based on its Java \glspl{dto} and \tool{Spring Boot}'s \gls{oas} extension.
We call this \gls{oas} type the \oasDefault{} \gls{oas}.
However, \glspl{oas} are often incomplete, which leads to ineffective \gls{rest} test generation tools~\citep{zhang:23a}.
This is also the case for \gls{crn}'s \oasDefault{} \gls{oas}, where medical variables are underconstrained which, consequently, lets the random input generator (informed by the \gls{oas}) try invalid variable values.
Our previous paper~\citep{laaber:23a} showed that, especially for validation rules, the four \evomaster{} tools struggle to generate tests that result in \rulePass{}, \ruleFail{}, or \ruleWarning{}, and most rules in most tests are being \ruleNotApplied{}.
Consequently, we augment the \oasDefault{} \gls{oas} by manually adding variable constraints to reflect the valid values as defined by a \gls{crn}-internal variable repository.
We call this \gls{oas} type the \oasStrict{} \gls{oas}.
Throughout this paper, we report results based on the \oasStrict{} \gls{oas} and compare to the \oasDefault{} \gls{oas} in \rqexecutedrules{}, \rqrulefrequency{}, and \rqmut{}.

\subsection{Rule Mutation Operators}
\label{sec:study:mutations}

\begin{table}[tbp]
    \centering
    \scriptsize
    \caption{Rule mutation operators based on \citet{isaku:24}}
    \label{tab:rule-mutation-operators}
    \begin{tabular}{llp{15.5em}}
        \toprule
        Name & ID & Description \\
        \midrule
        Alter Date & \mutAD{} & Alter the date by $\pm1$ year, $\pm1$ month, or $\pm1$ day. \\
        Negate Inequality & \mutNI{} & Change \code{=} to \code{!=} and vice versa. \\
        Reverse Inclusion & \mutRI{} & Change \code{in} condition to \code{notIn} and vice versa. \\
        Swap Comparison & \mutSComp{} & Swap \code{>} with \code{<=}, \code{<} with \code{>=}, and vice versa. \\
        Swap Connective & \mutSConn{} & Swap \code{and} with \code{or} and vice versa. \\
        Swap Startswith/Endswith & \mutSSE{} &  Swap \code{startswith} twith \code{endswith} and vice versa. \\
        Swap Subrule & \mutSSR{} & If \code{Rule A implies Rule B} then change it to \code{Rule B implies Rule A}. \\
        Swap Substring Indices & \mutSSI{} & If \code{substring(i,j)} then change it to \code{substring(j,i)}. \\
        \bottomrule
    \end{tabular}
\end{table}
 
To evaluate the tools with respect to their fault-finding ability, we employ artificial rule changes, i.e., rule mutations, in a mutation testing experiment in \rqmut{}.
For this, we adapt the \numMutationOperators{} rule mutation operators introduced in our previous work~\citep{isaku:24}, as shown in \cref{tab:rule-mutation-operators}.
In our previous work, we used the operators to generate a larger set of valid rules to use as test inputs, whereas in this paper, we employ the mutated rules as faulty \guri{} versions and execute the generated test suites (by the five tools) against it.
The mutation operators consider
\begin{inparaenum}
    \item altering dates (\mutAD{});
    \item changing set containment (\mutRI{});
    and
    \item swapping logical operators (\mutNI{}, \mutSComp{}, and \mutSConn{}), built-in functions (\mutSSE{}), subrule clauses (\mutSSR{}), and indices (\mutSSI{}).
\end{inparaenum}
We apply each operator one at a time on every rule at every possible location.
Therefore, each operator may be applied zero to multiple times per rule when there are no or multiple possible locations, e.g., a rule contains no or multiple \code{and} operators.
In version \version{v10}, applying the \numMutationOperators{} operators on the \num{70} validation and \num{43} aggregation rules leads to \numMutations{} rule mutations.

\subsection{Evaluation Metrics}
\label{sec:study:metrics}
To evaluate the effectiveness of the tools and answer the \glspl{rq}, we rely on the following evaluation metrics, which are the dependent variables of our study:
\begin{description}[leftmargin=0pt,labelindent=1.5em]
    \item[\rqcodecov{} (code coverage):]
    We rely on \textbf{line}, \textbf{branch}, and \textbf{method coverage} extracted with \tool{JaCoCo}\footnote{version \version{0.8.8} available at \url{https://www.jacoco.org/jacoco}}, similar to \citet{kim:22a}.
    \item[\rqfaults{} (code errors):]
    We use three types of \num{500} errors triggered by the tools, as defined by \citet{kim:22a}:
    \begin{inparaenum}
        \item \textbf{Unique Errors} are the number of errors grouped by their complete stack traces;
        \item \textbf{Unique Failure Points} are the number of occurrences of the same error, i.e., the first line of a stack trace;
        and
        \item \textbf{Unique Library Failure Points} are the number of errors that are unique failure points occurring in the library code.
    \end{inparaenum}
    \item[\rqexecutedrules{} (executed rules and results):]
    The first domain-specific metric is the number of (unique) executed rules per result type and rule type during a tool's execution.
    For example, a tool executes \num{70} validation rules with \rulePass{}, \num{5} with \ruleFail{}, \num{10} with \ruleWarning{}, and \num{70} with \ruleNotApplied{} of a total of \num{70} validation rules in \version{v10}.
    Note that each rule can have multiple result types during a tool's execution.
    \item[\rqrulefrequency{} (rule and result frequency):]
    The second domain-specific metric is the frequency of the result types per rule relative to the total number of rule executions during a tool's execution.
    For example, a tool executes rule \validationRule{01} \num{1}\% with \rulePass{}, \num{5}\% with \ruleFail{}, \num{0}\% with \ruleWarning{}, and \num{94}\% with \ruleNotApplied{}, totaling to \num{100}\% of all executions of this rule.
    Additionally, we compare the frequencies of the tools to production \guri{}, deployed at the \gls{crn}, to determine their similarity.
    \item[\rqmut{} (mutation testing):]
    The third domain-specific metric is inspired by the \gls{ms} of (code) mutation testing~\citep{papadakis:19}.
    Instead of code mutations, we rely on the rule mutations from \cref{sec:study:mutations}.
    The \gls{ms} is defined as $MS = \frac{M^{killed}}{M^{total}}$, where $M^{killed}$ is the number of \enquote{killed} mutants that lead to a failing test, and $M^{total}$ the total number of mutants, in our case \numMutations{}.
    We further differentiate between the \gls{ms} for validation and aggregation rules, which we denote as $MS^V$ and $MS^A$, respectively.
\end{description}

\subsection{Experiment Setup}
\label{sec:study:setup}

The experiment setup is concerned with the experiment execution settings and execution environment.

To deal with the stochastic nature of the \glspl{ea} underlying \evomaster{}, each tool is executed repeatedly for \num{30} repetitions~\citep{arcuri:11b}.
Each tool is executed for \num{1} hour for each repetition and version by following the practice of \citet{kim:22a}.
They also observed that for complex and constrained input parameters, source code coverage and thrown errors hardly increase after \num{10} minutes, which is also our case.
Beyond these settings, the experiments consider the tool parameterizations from \cref{sec:study:tools}.
The experiment further randomizes the order of the tools and versions in each repetition, with \gls{rmit}~\citep{abedi:17}, reducing potential confounding factors that stem from the execution environment or the execution order.

Mutation testing is known to be computationally expensive~\citep{papadakis:19}; in our context, we need to execute the \gls{rest} test suites against each of the \numMutations{} mutation.
A single test suite execution takes approximately \qty{10}{\second}.
For the five tools, \numVersions{} versions, \num{30} repetitions, two \gls{oas} types, this means that executing all test suites against all mutants takes approximately \num{791.3194444444} days when executed sequentially.
To keep a reasonable experiment runtime, we only execute the tools' test suites against \guri{} in (the most recent) version \version{v10} for a single repetition taking approximately \num{63.3055555556} days (sequentially).
We select the repetition with the median code coverage (from \rqcodecov{}).

We executed the experiments on the \gls{ex3} \gls{hpc} cluster\footnote{\url{https://www.ex3.simula.no}} hosted at the first author's institution, which uses Slurm \version{21.08.8-2} as its cluster management software.
The experiments were scheduled on nodes of the same type (using the \tool{rome16q} partition of \gls{ex3}).
The nodes have an AMD EPYC 7302P \num{16}-Core \glspl{cpu} @ \qty{3}{GHz} supporting simultaneous multithreading with \num{2} threads per core, run Ubuntu \version{22.04.5}, and have \qty{128}{GB} total memory.
The experiments were conducted in 2024.
The Java tools were exclusively built and ran with OpenJDK \version{17.0.11} built by Adoptium\footnote{\url{https://adoptium.net}}.

\subsection{Statistical Analyses}
\label{sec:study:stats}

Our previous conference paper~\citep{laaber:23b} did not report the results based on statistical tests, which we added in this paper.

We use hypothesis testing and effect size measures to compare the different observations.
An observation is a concrete metric value (see \cref{sec:study:metrics}) for a particular tool, \guri{} version, repetition, and \gls{oas} type.

Our experiments for \rqcodecov{} to \rqrulefrequency{} follow a complete block design, i.e., we compare five tools in \numVersions{} versions across \num{30} repeated measures (i.e., repetitions).
Hence, we employ the non-parametric Friedman test~\citep{friedman:37} and, from a statistical perspective, consider the tool-version pairs as individual subjects.
We state the null hypothesis H\textsubscript{0} that the subjects are the same,
and the alternative hypothesis H\textsubscript{1} that they are different.
Upon rejecting H\textsubscript{0}, we apply Nemenyi's post-hoc test~\citep{nemenyi:63} to identify which tool-version pairs are statistically different.

As effect size measure, we use the Vargha-Delaney \vda{}~\citep{vargha:00} to characterize the magnitude of the differences.
Two version-tool pairs are the same if \vda{}~$ = 0.5$.
If \vda{}~$> 0.5$, the first pair is larger than the second, otherwise if \vda{}~$< 0.5$.
The magnitude values are divided into four nominal categories, which rely on the scaled \vda{} defined
as $\hat{A}^{scaled}_{12} = (\hat{A}_{12} - 0.5) * 2$~\citep{hess:04}:
\enquote{negligible} ($|\hat{A}^{scaled}_{12}| < 0.147$),
\enquote{small} ($0.147 \leq |\hat{A}^{scaled}_{12}| < 0.33$),
\enquote{medium} ($0.33 \leq |\hat{A}^{scaled}_{12}| < 0.474$),
and
\enquote{large} ($|\hat{A}^{scaled}_{12}| \geq 0.474$).

We consider observations statistically significant different at a level $\alpha = 0.01$ and with a non-negligible effect size.
We control the false discovery rate with the Benjamini-Yekutieli procedure~\citep{benjamini:01}, which is considered a more powerful than, e.g., the Bonferroni or Holm corrections for family-wise Type I errors.

Note that we do not use statistical tests for \rqmut{} because we only have a single \gls{ms} value per tool (see \cref{sec:study:setup})

\subsection{Threats to Validity}
\label{sec:study:threats}

We classify threats into construct, internal, and external validity.

The biggest threat to the \textbf{construct validity} concerns the choice of the metrics for evaluating the tools' effectiveness.
We rely on metrics widely used in the \gls{api} test generation research, in particular, adapted by a recent publication~\citep{kim:22a}, to answer \rqcodecov{} and \rqfaults{}.
For \rqexecutedrules{} and \rqrulefrequency{}, we employ two sets of domain-specific metrics, i.e., the number of executed rules and their results in \rqexecutedrules{} and the frequency of result types per executed rule compared to the total number of results in \rqrulefrequency{}, which are of specific interest to the \gls{crn}.
For \rqmut{}, we adapt the well-known \gls{ms} for medical rules (instead of code mutations)~\citep{papadakis:19}.
Nevertheless, it is unclear whether these domain-specific metrics are correlated with \enquote{good} test cases for the domain experts or sufficiently targeted to assess test generation tools for the \gls{crn}.
Further investigation via dedicated empirical studies is required to answer this question. 

In terms of \textbf{internal validity}, a crucial threat is that \evomaster{} white-box requires manually creating a \gls{sut} driver.
Failure to do so concerning how \evomaster{} expects the driver to be implemented and \guri{} requires to be controlled could threaten the \evomasterwb{} results.
Similarly, an incorrect injection of the \tool{JaCoCo} code coverage agent, implementation of the analyses scripts, and adaptation of \guri{} to retrieve rule executions could alter the study's results and implications.
We thoroughly tested our implementations to validate the correct behavior to mitigate this threat.
Further internal validity threats relate to the experiment design include:
\begin{inparaenum}
    \item the number of repetitions (\num{30} in our study),
    \item the time budget for test generation (\num{1} hour),
    and
    \item the tool parameters (see \cref{sec:study:tools}).
\end{inparaenum}
These design decisions are based on previous research~\citep{arcuri:11b, kim:22a};
however, different experiment design decisions might lead to different results.
Finally, \evomaster{} relies on \tool{OpenAPI}\footnote{\url{https://www.openapis.org}} schema definitions to generate tests.
An incorrect schema definition potentially leads to suboptimal tests, which could impact the reported results.
Our experiments use the schema definitions generated by \tool{springdoc-openapi}\footnote{\url{https://springdoc.org}} (\oasDefault{} \gls{oas}), which \guri{} already employs, and refine these by manually adding variable constraints to form the \oasStrict{} \gls{oas}.
Considering the many parameters to configure, we try to build the base of our empirical study on the knowledge built by the existing literature to mitigate these internal threats. 

The primary \textbf{external validity threat} is related to the generalization is inherent to the study design: a case study on a single study subject, i.e., \gls{crn}['s] rule engine \guri{}.
All results are only valid in the context of our case study and are probably not transferable to other case studies.
Nevertheless, we provide implications and \enquote{more general} conclusions in the discussion section.
Moreover, our results are tightly coupled with the test generation tool(s), i.e., \evomaster{} in its four parameterizations and our \gls{crn}-specific extension \evoguri{} (see \cref{sec:study:tools}), and do not generalize to other \gls{rest} \gls{api} test generation tools.
Finally, we perform laboratory experiments on the research \gls{hpc} cluster \gls{ex3} with a standalone version of \guri{} and not a field experiment against the real-world \guri{} (or the testing environment) hosted at the \gls{crn}.
Consequently, our results might not generalize to the real system.
The extraction of the real-world \guri{} into a standalone version for the empirical study, was, however, done by the \gls{crn} developers, which should reduce this threat.
 
\section{Results}
\label{sec:results}
This section presents the results for each \gls{rq}.

\subsection{\rqcodecov{}: Code Coverage}
\label{sec:results:code-coverage}

\begin{table*}
    \centering
    \scriptsize
    \caption{
    Source code coverage per tool across all the versions and repetitions, each compared to the other tools in each version and between the \gls{oas} types.
    }
    \label{tab:coverages}
    
    \begin{tabular}{lllrrrlrrr}
\toprule
Coverage & Tool & Mean$\pm$SD & \nrcols{3}{Tool Comp.} & \nrcols{4}{\Gls{oas} Comp.} \\
\cmidrule(l{1pt}r{1pt}){4-6}
\cmidrule(l{1pt}r{1pt}){7-10}
 &  &  & $<$ & $=$ & $>$ & Diff. Mean$\pm$SD & $<$ & $=$ & $>$ \\
\midrule
\covLine{} & \evoguriRandom{} & \num{20.9866666666667}\%$\pm$\num{0.114889312249309} & \num{0} & \num{25} & \num{15} & \num{0.983333333333333}~\gls{pp}$\pm$\num{0.0283278861866273} & \num{0} & \num{0} & \num{10} \\
 & \evomasterbb{} & \num{20.9966666666667}\%$\pm$\num{0.0577350269189624} & \num{0} & \num{25} & \num{15} & \num{1}~\gls{pp}$\pm$\num{0.0157134840263685} & \num{0} & \num{0} & \num{10} \\
 & \evomasterwbmio{} & \num{20.0533333333333}\%$\pm$\num{0.564106618005096} & \num{26} & \num{14} & \num{0} & \num{4.72}~\gls{pp}$\pm$\num{0.870532687780205} & \num{0} & \num{0} & \num{10} \\
 & \evomasterwbmosa{} & \num{20.4233333333333}\%$\pm$\num{0.494912804950429} & \num{13} & \num{27} & \num{0} & \num{2.24666666666667}~\gls{pp}$\pm$\num{0.423448511598881} & \num{0} & \num{0} & \num{10} \\
 & \evomasterwbwts{} & \num{20.7666666666667}\%$\pm$\num{0.42365927286816} & \num{0} & \num{31} & \num{9} & \num{1.90333333333333}~\gls{pp}$\pm$\num{0.237502436634674} & \num{0} & \num{0} & \num{10} \\
\midrule
\covBranch{} & \evoguriRandom{} & \num{16.3966666666667}\%$\pm$\num{0.490023092305042} & \num{0} & \num{24} & \num{16} & \num{1.81666666666667}~\gls{pp}$\pm$\num{0.160439211979929} & \num{0} & \num{0} & \num{10} \\
 & \evomasterbb{} & \num{16.4}\%$\pm$\num{0.490716492065718} & \num{0} & \num{24} & \num{16} & \num{1.59666666666667}~\gls{pp}$\pm$\num{0.368999314128306} & \num{0} & \num{0} & \num{10} \\
 & \evomasterwbmio{} & \num{15.14}\%$\pm$\num{0.664735217736352} & \num{30} & \num{10} & \num{0} & \num{5.89}~\gls{pp}$\pm$\num{1.01191055087046} & \num{0} & \num{0} & \num{10} \\
 & \evomasterwbmosa{} & \num{15.73}\%$\pm$\num{0.480836784551604} & \num{9} & \num{30} & \num{1} & \num{3.33}~\gls{pp}$\pm$\num{0.302438647997357} & \num{0} & \num{0} & \num{10} \\
 & \evomasterwbwts{} & \num{16.1233333333333}\%$\pm$\num{0.402486696290937} & \num{4} & \num{26} & \num{10} & \num{3}~\gls{pp}$\pm$\num{0.119670329047433} & \num{0} & \num{0} & \num{10} \\
\midrule
\covMethod{} & \evoguriRandom{} & \num{21}\%$\pm$\num{0} & \num{0} & \num{35} & \num{5} & \num{0.91}~\gls{pp}$\pm$\num{0.121766828380422} & \num{0} & \num{0} & \num{10} \\
 & \evomasterbb{} & \num{21}\%$\pm$\num{0} & \num{0} & \num{35} & \num{5} & \num{0.893333333333333}~\gls{pp}$\pm$\num{0.151372323252679} & \num{0} & \num{0} & \num{10} \\
 & \evomasterwbmio{} & \num{20.54}\%$\pm$\num{0.505885098062271} & \num{15} & \num{25} & \num{0} & \num{3.33666666666667}~\gls{pp}$\pm$\num{0.617931635517101} & \num{0} & \num{0} & \num{10} \\
 & \evomasterwbmosa{} & \num{20.7866666666667}\%$\pm$\num{0.410345586740509} & \num{0} & \num{40} & \num{0} & \num{1.57}~\gls{pp}$\pm$\num{0.223579190268207} & \num{0} & \num{0} & \num{10} \\
 & \evomasterwbwts{} & \num{20.9833333333333}\%$\pm$\num{0.12823299585641} & \num{0} & \num{35} & \num{5} & \num{1.31666666666667}~\gls{pp}$\pm$\num{0.0849836585598796} & \num{0} & \num{0} & \num{10} \\
\bottomrule
\end{tabular}
 \end{table*}
 
We study the code coverage achieved by the tools in terms of line, branch, and method coverages.
\Cref{tab:coverages} depicts the coverage results.
We refrain from reporting coverages for each version, as the source code does not change between versions; only the rules change (see \cref{sec:study:subject}).
Each row reports the coverage in percent as the arithmetic mean and standard deviation (\enquote{Mean$\pm$SD}) across all the versions and repetitions, as well as the coverage difference in \gls{pp} for each version when using the \oasStrict{} and \oasDefault{} \glspl{oas} (\enquote{Diff. Mean$\pm$SD}).
In addition, we provide the results of the statistical tests.
\begin{inparaenum}
    \item The \enquote{Tool Comp.} columns count how often the tool is worse (\enquote{$<$}), equal (\enquote{$=$}), and better (\enquote{$>$}) than the other tools.
    Note that we compare each tool in each version against the other tools in the same version and, hence, we report \num{40} comparisons (\num{4} other tools in \num{10} versions).
    \item The \enquote{\gls{oas} Comp.} columns count how often the same tool is worse (\enquote{$<$}), equal (\enquote{$=$}), and better (\enquote{$>$}) when using the \oasStrict{} or \oasDefault{} \glspl{oas} per version (\num{10} comparisons).
\end{inparaenum}

We observe that all tools perform similarly, i.e., approximately \num{21}\% line coverage, \num{16}\% branch coverage, and \num{21}\% method coverage.
Note that the relatively low (absolute) coverage values must be considered with caution because we generated tests for (only) \num{2} of \num{32} \gls{rest} endpoints (see \cref{sec:study:subject}), as only these two handle the medical rules.
However, the code coverage is still of interest as a relative comparison among the tools.
Regarding it, the tools perform similarly, which is a different observation as obtained by \citet{kim:22a}, where \evomasterwbmio{} achieves a higher code coverage than \evomasterbb{} by \num{7.35}~\gls{pp} (line), \num{7.87}~\gls{pp} (branch), and \num{5.69}~\gls{pp} (method).
In addition, \citet{kim:22a} report that the line and method coverages are similarly higher than the branch coverage, which aligns with our findings.
A potential reason for the low coverage values is that all the tools reach the maximum code coverage achievable through the two \gls{rest} \glspl{api} used in our experiment.

In terms of statistical tests, we observe that, although the mean differences among the tools are minuscule, there are statistical differences:
\evoguri{} and \evomasterbb{} perform equally well by being better than \num{15} tool-version pairs for line coverage, \num{16} for branch coverage, and \num{5} for method coverage, as well as never worse.
This shows that black-box approaches are superior to white-box approaches in our setting and suggests that a simple random sampling leads to (slightly) better results than trying to employ a sophisticated search algorithm.

When comparing the two \gls{oas} types, we observe that all the tools for all three coverage types perform better when using the \oasStrict{} \gls{oas}, showing a mean improvement between \num{0.98} and \num{5.89} \gls{pp},
Especially, the \evomasterwb{} tools benefit from the \oasStrict{} \gls{oas}, as they spend less time evolving an already invalid request.
This result is in line with \citet{kim:22a} who also found that test generation tools often suffer from bad performance due to sampling invalid input data.

\summarybox{\rqcodecov{} Summary}{
All the tools achieve a similar line, branch, and method coverage, with a slight edge for \evoguri{} and \evomasterbb{}, which are statistically equal to or better than all the other tools.
In the context of the \gls{crn}, opting for a simpler test generation tool, i.e., \evoguri{} or \evomasterbb{}, is preferred over a more complex tool, i.e., any \evomasterwb{}, to cover more source code.
}

\subsection{\rqfaults{}: Code Errors}
\label{sec:results:errors}

The second \gls{rq} studies the thrown errors related to source code by the tools in terms of unique \num{500} errors, unique failure points, and unique library failure points.
\Cref{tab:errors} shows the results for each tool across all the versions and repetitions.
Similar to \rqcodecov{}, we refrain from reporting errors for each version.
For each error type (\enquote{Error Type}, see \cref{sec:study:metrics}), the table depicts four categories (\enquote{Error Cat.}):
\begin{inparaenum}
    \item the total number of errors (\enquote{All}, higher is better);
    \item errors that are due to the tools' intervention, e.g., the attached Java agent (\enquote{Tool}, lower is better);
    \item \gls{io} errors that occur during the test generation (\enquote{\gls{io}}, lower is better);
    and
    \item remaining errors that are not attributed to the Tool and \gls{io} categories but actually due to errors thrown by the application, i.e., \guri{} (\enquote{Remaining}, higher is better).
\end{inparaenum}
In addition, the table reports the results of the statistical tests for each tool and version comparing
\begin{inparaenum}
    \item to every other tool and version (\enquote{Tool Comp.})
    and
    \item the \oasStrict{} and \oasDefault{} \glspl{oas} (\enquote{\gls{oas} Comp.}),
\end{inparaenum}
similar to \rqcodecov{}.

\begin{table*}
    \centering
    \scriptsize
    \caption{
    500 errors per tool across all the versions and repetitions, each compared to the other tools in each version and between the \gls{oas} types.
    }
    \label{tab:errors}
    
    \begin{tabular}{llllrrrlrrr}
\toprule
Tool & \nrcols{2}{Error} & Mean$\pm$SD & \nrcols{3}{Tool Comp.} & \nrcols{4}{\Gls{oas} Comp.} \\
\cmidrule(l{1pt}r{1pt}){2-3}
\cmidrule(l{1pt}r{1pt}){5-7}
\cmidrule(l{1pt}r{1pt}){8-11}
 & Type & Cat. &  & $<$ & $=$ & $>$ & Diff. Mean$\pm$SD & $<$ & $=$ & $>$ \\
\midrule
\evoguriRandom{} & Unique Errors & All & \num{4.34333333333333}$\pm$\num{1.54057352247447} & \num{10} & \num{28} & \num{2} & \num{-0.0566666666666666}$\pm$\num{0.944908382782083} & \num{1} & \num{9} & \num{0} \\
 &  & Tool & \num{0}$\pm$\num{0} & \num{0} & \num{10} & \num{30} & \num{0}$\pm$\num{0} & \num{0} & \num{10} & \num{0} \\
 &  & \Gls{io} & \num{1.82333333333333}$\pm$\num{1.23715698826014} & \num{12} & \num{28} & \num{0} & \num{-1.39}$\pm$\num{0.826273801509965} & \num{0} & \num{2} & \num{8} \\
 &  & Remaining & \num{2.52}$\pm$\num{0.835820146757514} & \num{0} & \num{27} & \num{13} & \num{1.33333333333333}$\pm$\num{0.329608821648696} & \num{0} & \num{0} & \num{10} \\
 & Unique Failure Points & All & \num{1.99333333333333}$\pm$\num{0.141500164466461} & \num{7} & \num{33} & \num{0} & \num{-0.153333333333333}$\pm$\num{0.0849109918868568} & \num{0} & \num{10} & \num{0} \\
 &  & Tool & \num{0}$\pm$\num{0} & \num{0} & \num{10} & \num{30} & \num{0}$\pm$\num{0} & \num{0} & \num{10} & \num{0} \\
 &  & \Gls{io} & \num{0.77}$\pm$\num{0.421535654117547} & \num{0} & \num{39} & \num{1} & \num{-0.213333333333333}$\pm$\num{0.261618891604648} & \num{0} & \num{9} & \num{1} \\
 &  & Remaining & \num{1.22333333333333}$\pm$\num{0.417175831599168} & \num{0} & \num{26} & \num{14} & \num{0.06}$\pm$\num{0.291356978445495} & \num{0} & \num{9} & \num{1} \\
 & Unique Library Failure Points &  & \num{1}$\pm$\num{0} & \num{0} & \num{30} & \num{10} & \num{0.976666666666667}$\pm$\num{0.0274424200782855} & \num{0} & \num{0} & \num{10} \\
\midrule
\evomasterbb{} & Unique Errors & All & \num{4.56666666666667}$\pm$\num{1.82299201548369} & \num{9} & \num{27} & \num{4} & \num{0.266666666666667}$\pm$\num{1.07324361253718} & \num{1} & \num{9} & \num{0} \\
 &  & Tool & \num{0}$\pm$\num{0} & \num{0} & \num{10} & \num{30} & \num{0}$\pm$\num{0} & \num{0} & \num{10} & \num{0} \\
 &  & \Gls{io} & \num{1.98}$\pm$\num{1.38539925785996} & \num{12} & \num{28} & \num{0} & \num{-1.13}$\pm$\num{0.836873959859762} & \num{0} & \num{5} & \num{5} \\
 &  & Remaining & \num{2.58666666666667}$\pm$\num{0.88227940323543} & \num{0} & \num{25} & \num{15} & \num{1.39666666666667}$\pm$\num{0.332015915132445} & \num{0} & \num{0} & \num{10} \\
 & Unique Failure Points & All & \num{2.01666666666667}$\pm$\num{0.288675134594813} & \num{7} & \num{33} & \num{0} & \num{-0.0966666666666667}$\pm$\num{0.103577966101524} & \num{0} & \num{10} & \num{0} \\
 &  & Tool & \num{0}$\pm$\num{0} & \num{0} & \num{10} & \num{30} & \num{0}$\pm$\num{0} & \num{0} & \num{10} & \num{0} \\
 &  & \Gls{io} & \num{0.726666666666667}$\pm$\num{0.446415177375119} & \num{0} & \num{40} & \num{0} & \num{-0.223333333333333}$\pm$\num{0.257744250693264} & \num{0} & \num{9} & \num{1} \\
 &  & Remaining & \num{1.29}$\pm$\num{0.503399812512339} & \num{0} & \num{23} & \num{17} & \num{0.126666666666667}$\pm$\num{0.33065591380366} & \num{0} & \num{9} & \num{1} \\
 & Unique Library Failure Points &  & \num{1}$\pm$\num{0} & \num{0} & \num{30} & \num{10} & \num{0.983333333333333}$\pm$\num{0.0235702260395516} & \num{0} & \num{0} & \num{10} \\
\midrule
\evomasterwbmio{} & Unique Errors & All & \num{4.73333333333333}$\pm$\num{3.32350837448518} & \num{3} & \num{32} & \num{5} & \num{1.94333333333333}$\pm$\num{0.805084153258006} & \num{0} & \num{10} & \num{0} \\
 &  & Tool & \num{2.71666666666667}$\pm$\num{1.97214436596892} & \num{20} & \num{20} & \num{0} & \num{1.34}$\pm$\num{0.57236783476328} & \num{0} & \num{10} & \num{0} \\
 &  & \Gls{io} & \num{1.10666666666667}$\pm$\num{0.915468457792475} & \num{0} & \num{32} & \num{8} & \num{0.736666666666667}$\pm$\num{0.257456696586599} & \num{0} & \num{10} & \num{0} \\
 &  & Remaining & \num{0.91}$\pm$\num{1.38623186948887} & \num{18} & \num{22} & \num{0} & \num{-0.133333333333333}$\pm$\num{0.301027048536535} & \num{1} & \num{9} & \num{0} \\
 & Unique Failure Points & All & \num{1.96666666666667}$\pm$\num{1.27971917209218} & \num{3} & \num{37} & \num{0} & \num{0.243333333333333}$\pm$\num{0.419008235282516} & \num{0} & \num{10} & \num{0} \\
 &  & Tool & \num{0.76}$\pm$\num{0.479408777474644} & \num{20} & \num{20} & \num{0} & \num{0.36}$\pm$\num{0.150554530541816} & \num{0} & \num{10} & \num{0} \\
 &  & \Gls{io} & \num{0.863333333333333}$\pm$\num{0.65784859311275} & \num{1} & \num{39} & \num{0} & \num{0.563333333333333}$\pm$\num{0.219117196219312} & \num{1} & \num{9} & \num{0} \\
 &  & Remaining & \num{0.343333333333333}$\pm$\num{0.475614957201659} & \num{21} & \num{19} & \num{0} & \num{-0.68}$\pm$\num{0.115683689642848} & \num{9} & \num{1} & \num{0} \\
 & Unique Library Failure Points &  & \num{0.333333333333333}$\pm$\num{0.472192164649887} & \num{23} & \num{17} & \num{0} & \num{0.333333333333333}$\pm$\num{0.111111111111111} & \num{0} & \num{10} & \num{0} \\
\midrule
\evomasterwbmosa{} & Unique Errors & All & \num{4.17666666666667}$\pm$\num{2.01635177249831} & \num{6} & \num{32} & \num{2} & \num{0.426666666666667}$\pm$\num{0.59271295074033} & \num{0} & \num{10} & \num{0} \\
 &  & Tool & \num{2.09666666666667}$\pm$\num{1.08223209614295} & \num{20} & \num{20} & \num{0} & \num{-0.4}$\pm$\num{0.268971520820227} & \num{0} & \num{10} & \num{0} \\
 &  & \Gls{io} & \num{0.72}$\pm$\num{0.776924732605277} & \num{0} & \num{32} & \num{8} & \num{0.503333333333333}$\pm$\num{0.186884793190687} & \num{2} & \num{8} & \num{0} \\
 &  & Remaining & \num{1.36}$\pm$\num{1.19799163039159} & \num{9} & \num{31} & \num{0} & \num{0.323333333333333}$\pm$\num{0.20370033667251} & \num{0} & \num{10} & \num{0} \\
 & Unique Failure Points & All & \num{2.11}$\pm$\num{0.990540543623468} & \num{2} & \num{38} & \num{0} & \num{-0.0733333333333334}$\pm$\num{0.337309617084627} & \num{0} & \num{10} & \num{0} \\
 &  & Tool & \num{0.87}$\pm$\num{0.374478455605091} & \num{20} & \num{20} & \num{0} & \num{-0.0766666666666667}$\pm$\num{0.0667591950479991} & \num{0} & \num{10} & \num{0} \\
 &  & \Gls{io} & \num{0.63}$\pm$\num{0.638586440617453} & \num{0} & \num{40} & \num{0} & \num{0.416666666666667}$\pm$\num{0.233201020686837} & \num{2} & \num{8} & \num{0} \\
 &  & Remaining & \num{0.61}$\pm$\num{0.488564890443341} & \num{9} & \num{31} & \num{0} & \num{-0.413333333333333}$\pm$\num{0.0834443704689715} & \num{0} & \num{10} & \num{0} \\
 & Unique Library Failure Points &  & \num{0.61}$\pm$\num{0.488564890443341} & \num{2} & \num{38} & \num{0} & \num{0.61}$\pm$\num{0.0861379886369525} & \num{0} & \num{3} & \num{7} \\
\midrule
\evomasterwbwts{} & Unique Errors & All & \num{5.64333333333333}$\pm$\num{1.64879267338912} & \num{0} & \num{25} & \num{15} & \num{1.60333333333333}$\pm$\num{0.484181886036056} & \num{0} & \num{8} & \num{2} \\
 &  & Tool & \num{2.64666666666667}$\pm$\num{0.870448826422489} & \num{20} & \num{20} & \num{0} & \num{-0.2}$\pm$\num{0.247954595604682} & \num{0} & \num{10} & \num{0} \\
 &  & \Gls{io} & \num{1}$\pm$\num{0.784464540552736} & \num{0} & \num{32} & \num{8} & \num{0.81}$\pm$\num{0.202484567313166} & \num{5} & \num{5} & \num{0} \\
 &  & Remaining & \num{1.99666666666667}$\pm$\num{0.912559461876554} & \num{3} & \num{35} & \num{2} & \num{0.993333333333333}$\pm$\num{0.206559111797729} & \num{0} & \num{2} & \num{8} \\
 & Unique Failure Points & All & \num{2.79333333333333}$\pm$\num{0.696831497572348} & \num{0} & \num{21} & \num{19} & \num{0.586666666666667}$\pm$\num{0.225092573548455} & \num{0} & \num{9} & \num{1} \\
 &  & Tool & \num{1.02333333333333}$\pm$\num{0.190376028769351} & \num{20} & \num{20} & \num{0} & \num{0.00666666666666669}$\pm$\num{0.040975753143524} & \num{0} & \num{10} & \num{0} \\
 &  & \Gls{io} & \num{0.87}$\pm$\num{0.611839706203315} & \num{0} & \num{40} & \num{0} & \num{0.683333333333333}$\pm$\num{0.171593835683117} & \num{5} & \num{5} & \num{0} \\
 &  & Remaining & \num{0.9}$\pm$\num{0.311432121906499} & \num{4} & \num{33} & \num{3} & \num{-0.103333333333333}$\pm$\num{0.0776983721646537} & \num{0} & \num{10} & \num{0} \\
 & Unique Library Failure Points &  & \num{0.896666666666667}$\pm$\num{0.304902345607155} & \num{0} & \num{35} & \num{5} & \num{0.896666666666667}$\pm$\num{0.0760928607478397} & \num{0} & \num{0} & \num{10} \\
\bottomrule
\end{tabular}
 \end{table*}
 
\subsubsection{Unique Errors}
We observe that the different tools trigger between \num{4.18} and \num{5.64} errors (\enquote{All});
however, on a closer inspection, we notice that the three \evomasterwb{} tools experience a high number of tool-related errors, i.e., the coverage agent attached to the \gls{jvm} throws a \code{KillSwitch} exception when concurrent threads are running after the test generation has finished.
This scenario inflates the \enquote{All} errors.
Moreover, we can see that all the tools suffer from a varying degree of \gls{io} errors, most often due to broken \gls{io} pipe exceptions.
However, \evoguri{} and \evomasterbb{} suffer more from \gls{io} errors than \evomasterwb{}.
Finally, the number of \enquote{Remaining} unique errors varies the five tools between \num{0.91} and \num{2.59}.
More pronounced than for \rqcodecov{}, \evoguri{} and \evomasterbb{} outperform the three \evomasterwb{} tools in the \enquote{Remaining} category, which is most important as it depicts real errors.
A potential reason is that trying more randomly sampled inputs leads to finding more unique errors consistently.

The statistical tests reveal that \evoguri{} is in \num{27} cases equal and in \num{13} cases better than the other tool-version pairs in finding errors (i.e., \enquote{Remaining} category).
Similarly, \evomasterbb{} is equal in \num{25} and better in \num{15} cases.
When using the \oasStrict{} \gls{oas} instead of the \oasDefault{} \gls{oas}, \evoguri{}, \evomasterbb{}, and \evomasterwbwts{} find more errors, while \evomasterwbmio{} and \evomasterwbmosa{} or largely unaffected.

\subsubsection{Unique Failure Points}
We observe a similar trend:
\enquote{All} errors are inflated by tool-related and \gls{io} errors, leaving between \num{0.34} and \num{1.29} \enquote{Remaining} unique failure points, where \evoguri{} and \evomasterbb{} outperform \evomasterwb{}.
In most cases, this failure point is caused by a date parsing exception, e.g., when the diagnosis date of a cancer message is malformed.
All the tools find this failure point, however, \evoguri{} and \evomasterbb{} find it more consistently.

The statistical tests yield similar results as for the unique errors.
\evoguri{} and \evomasterbb{} are either equal or better than the remaining tools, except for \gls{io} errors.
In terms of \glspl{oas} types, \oasStrict{} leads to (mostly) the same number of unique failure points across all four tools.

\subsubsection{Unique Library Failure Points}
Although seldom, all four tools trigger a unique library failure point (i.e., a unique failure point where the cause is located in \guri{}'s source code, see \cref{sec:study:metrics}).
Upon closer inspection, we identify \num{1} \guri{} rule parsing error on specific inputs.
Similar to the unique failure points, \evoguri{} and \evomasterbb{} consistently find this failure point across all the versions.
The \evomasterwb{} tools find it to a varying degree in \num{0.33} (\evomasterwbmio{}), \num{0.61} (\evomasterwbmosa{}), and \num{0.9} (\evomasterwbwts{}) of the versions.

The statistical tests are in line with the unique errors and unique failure points, \evoguri{} and \evomasterbb{} are equal to \num{30} and better than \num{10}, and \evomasterwbwts{} is equal to \num{35} and better than \num{5} tool-version pairs.
Using \oasStrict{} \gls{oas} improves the failure point discovery for \evoguri{}, \evomasterbb{}, and \evomasterwbwts{} in all versions, for \evomasterwbmosa{} in \num{7} versions, and has no effect on \evomasterwbmio{}.

\summarybox{\rqfaults{} Summary}{
    \evoguri{} and \evomasterbb{} are tied for the most effective tool in finding errors, failure points, and library failure points, outperforming the \evomasterwb{} tools.
    Similar to \rqcodecov{}, this suggests employing one of these two at the \gls{crn} for their error-revealing capabilities.
}

\subsection{\rqexecutedrules{}: Executed Rules and Results}
\label{sec:results:executed-rules}

\begin{figure*}[tbp]
    \centering
    \includegraphics[width=\textwidth]{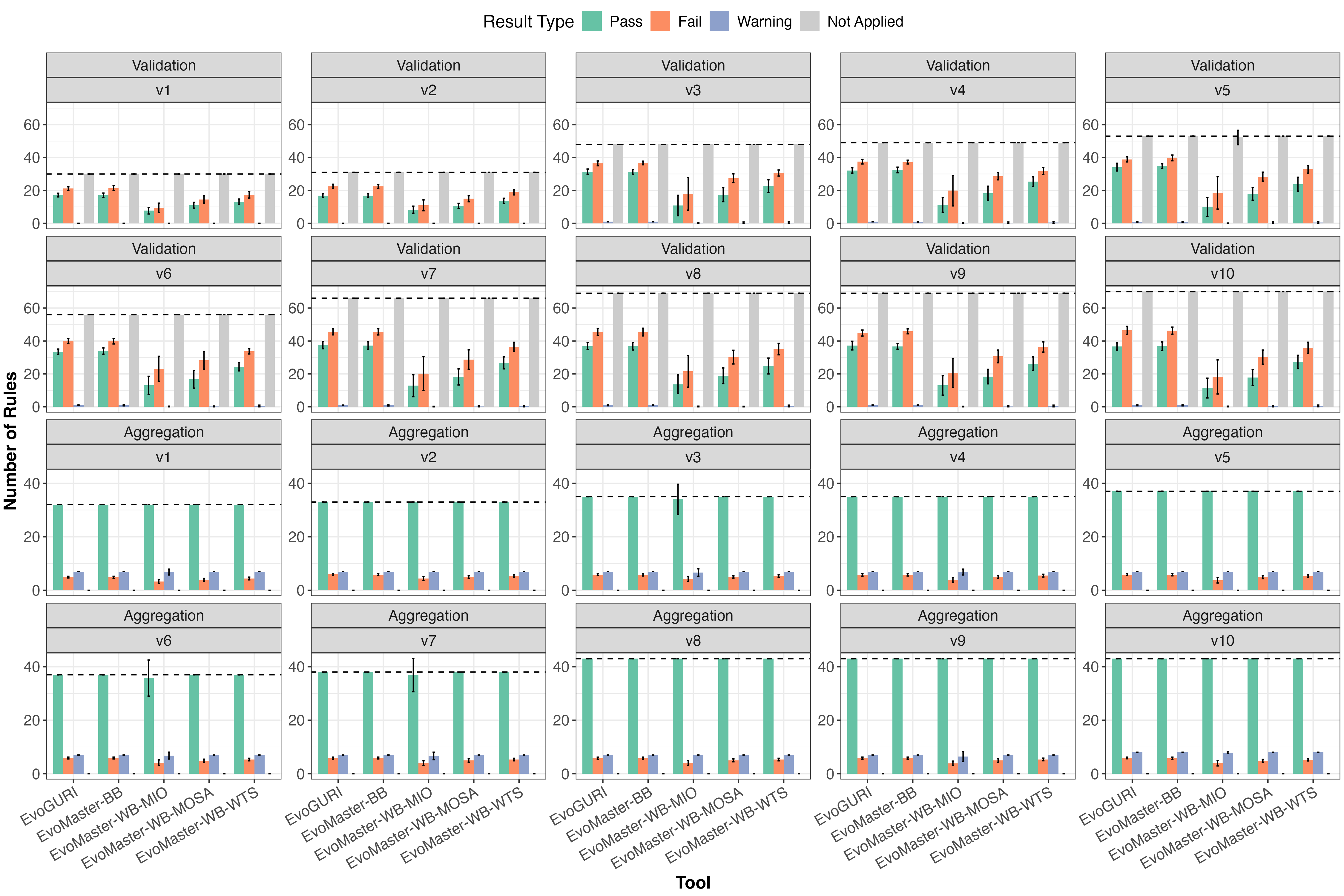}
    \caption{
    Number of executed rules and results per tool and version across all the repetitions.
    The bars are arithmetic means, the error bars are standard deviations, and the dashed line depicts the number of total rules of a version.
    }
    \label{fig:executed-rules}
\end{figure*}
 
\begin{table*}
    \centering
    \scriptsize
    \caption{
    Executed rules and results per tool across all the versions and repetitions, each compared to the other tools in each version and between the \gls{oas} types.
    }
    \label{tab:executed-rules:stats}
    
    \begin{tabular}{lllllrrrlrrr}
\toprule
Rule Type & \# Rules & Tool & Result Type & Mean$\pm$SD & \nrcols{3}{Tool Comp.} & \nrcols{4}{\Gls{oas} Comp.} \\
\cmidrule(l{1pt}r{1pt}){6-8}
\cmidrule(l{1pt}r{1pt}){9-12}
 &  &  &  &  & $<$ & $=$ & $>$ & Diff. Mean$\pm$SD & $<$ & $=$ & $>$ \\
\midrule
Validation & \num{54.1}$\pm$\num{14.2395886979795} & \evoguriRandom{} & \rulePass{} & \num{31.3566666666667}$\pm$\num{7.69242268587402} & \num{0} & \num{10} & \num{30} & \num{26.95}$\pm$\num{7.51964094887515} & \num{0} & \num{0} & \num{10} \\
 &  &  & \ruleFail{} & \num{37.8633333333333}$\pm$\num{8.90631219296089} & \num{0} & \num{10} & \num{30} & \num{32.7266666666667}$\pm$\num{8.92033050375917} & \num{0} & \num{0} & \num{10} \\
 &  &  & \ruleWarning{} & \num{0.776666666666667}$\pm$\num{0.41717583159917} & \num{0} & \num{20} & \num{20} & \num{0.776666666666667}$\pm$\num{0.410073766324012} & \num{0} & \num{2} & \num{8} \\
 &  &  & \ruleNotApplied{} & \num{54.1}$\pm$\num{14.2395886979795} & \num{0} & \num{40} & \num{0} & \num{0}$\pm$\num{0} & \num{0} & \num{10} & \num{0} \\
\cmidrule{3-12}
 &  & \evomasterbb{} & \rulePass{} & \num{31.4}$\pm$\num{7.69267557648726} & \num{0} & \num{10} & \num{30} & \num{26.9933333333333}$\pm$\num{7.56448491959403} & \num{0} & \num{0} & \num{10} \\
 &  &  & \ruleFail{} & \num{38.06}$\pm$\num{8.91673110761365} & \num{0} & \num{10} & \num{30} & \num{32.64}$\pm$\num{8.84159362206711} & \num{0} & \num{0} & \num{10} \\
 &  &  & \ruleWarning{} & \num{0.75}$\pm$\num{0.43373619897468} & \num{0} & \num{23} & \num{17} & \num{0.75}$\pm$\num{0.397290203972914} & \num{0} & \num{2} & \num{8} \\
 &  &  & \ruleNotApplied{} & \num{54.1}$\pm$\num{14.2395886979795} & \num{0} & \num{40} & \num{0} & \num{0}$\pm$\num{0} & \num{0} & \num{10} & \num{0} \\
\cmidrule{3-12}
 &  & \evomasterwbmio{} & \rulePass{} & \num{11.2}$\pm$\num{5.53674257977096} & \num{30} & \num{10} & \num{0} & \num{10.22}$\pm$\num{2.10673230089974} & \num{0} & \num{0} & \num{10} \\
 &  &  & \ruleFail{} & \num{18.03}$\pm$\num{9.47143743527354} & \num{29} & \num{11} & \num{0} & \num{16.2166666666667}$\pm$\num{4.78103893869662} & \num{0} & \num{0} & \num{10} \\
 &  &  & \ruleWarning{} & \num{0.0733333333333333}$\pm$\num{0.261118424591396} & \num{16} & \num{24} & \num{0} & \num{0.0733333333333333}$\pm$\num{0.0491909858248415} & \num{0} & \num{10} & \num{0} \\
 &  &  & \ruleNotApplied{} & \num{54.02}$\pm$\num{14.3130176733066} & \num{0} & \num{40} & \num{0} & \num{17.4433333333333}$\pm$\num{8.03507818763959} & \num{0} & \num{0} & \num{10} \\
\cmidrule{3-12}
 &  & \evomasterwbmosa{} & \rulePass{} & \num{16.5233333333333}$\pm$\num{5.03859242258112} & \num{20} & \num{20} & \num{0} & \num{15.8}$\pm$\num{3.0488613539411} & \num{0} & \num{0} & \num{10} \\
 &  &  & \ruleFail{} & \num{26.2066666666667}$\pm$\num{6.88927882762568} & \num{20} & \num{20} & \num{0} & \num{25.17}$\pm$\num{6.099290588176} & \num{0} & \num{0} & \num{10} \\
 &  &  & \ruleWarning{} & \num{0.196666666666667}$\pm$\num{0.398142285899938} & \num{16} & \num{24} & \num{0} & \num{0.196666666666667}$\pm$\num{0.140061714792506} & \num{0} & \num{10} & \num{0} \\
 &  &  & \ruleNotApplied{} & \num{54.1}$\pm$\num{14.2395886979795} & \num{0} & \num{40} & \num{0} & \num{7.93}$\pm$\num{4.7138998559422} & \num{0} & \num{9} & \num{1} \\
\cmidrule{3-12}
 &  & \evomasterwbwts{} & \rulePass{} & \num{22.79}$\pm$\num{5.98151276712484} & \num{20} & \num{10} & \num{10} & \num{22.1833333333333}$\pm$\num{5.11452180886281} & \num{0} & \num{0} & \num{10} \\
 &  &  & \ruleFail{} & \num{30.8833333333333}$\pm$\num{7.11466446082945} & \num{20} & \num{11} & \num{9} & \num{30.0266666666667}$\pm$\num{7.0274734233613} & \num{0} & \num{0} & \num{10} \\
 &  &  & \ruleWarning{} & \num{0.373333333333333}$\pm$\num{0.484497694119983} & \num{5} & \num{35} & \num{0} & \num{0.373333333333333}$\pm$\num{0.201720990540626} & \num{0} & \num{5} & \num{5} \\
 &  &  & \ruleNotApplied{} & \num{54.1}$\pm$\num{14.2395886979795} & \num{0} & \num{40} & \num{0} & \num{2.15666666666667}$\pm$\num{1.94454285465254} & \num{0} & \num{10} & \num{0} \\
\midrule
Aggregation & \num{37.6}$\pm$\num{3.93594191979335} & \evoguriRandom{} & \rulePass{} & \num{37.6}$\pm$\num{3.93594191979335} & \num{0} & \num{40} & \num{0} & \num{0}$\pm$\num{0} & \num{0} & \num{10} & \num{0} \\
 &  &  & \ruleFail{} & \num{5.74333333333333}$\pm$\num{0.459884625151125} & \num{0} & \num{19} & \num{21} & \num{2.27333333333333}$\pm$\num{0.105174751699549} & \num{0} & \num{0} & \num{10} \\
 &  &  & \ruleWarning{} & \num{7.1}$\pm$\num{0.300501253482382} & \num{0} & \num{40} & \num{0} & \num{0}$\pm$\num{0} & \num{0} & \num{10} & \num{0} \\
 &  &  & \ruleNotApplied{} & \num{0}$\pm$\num{0} & \num{0} & \num{40} & \num{0} & \num{0}$\pm$\num{0} & \num{0} & \num{10} & \num{0} \\
\cmidrule{3-12}
 &  & \evomasterbb{} & \rulePass{} & \num{37.6}$\pm$\num{3.93594191979335} & \num{0} & \num{40} & \num{0} & \num{0}$\pm$\num{0} & \num{0} & \num{10} & \num{0} \\
 &  &  & \ruleFail{} & \num{5.71666666666667}$\pm$\num{0.487010767005709} & \num{0} & \num{20} & \num{20} & \num{2.30666666666667}$\pm$\num{0.100369687027877} & \num{0} & \num{0} & \num{10} \\
 &  &  & \ruleWarning{} & \num{7.1}$\pm$\num{0.300501253482382} & \num{0} & \num{40} & \num{0} & \num{0}$\pm$\num{0} & \num{0} & \num{10} & \num{0} \\
 &  &  & \ruleNotApplied{} & \num{0}$\pm$\num{0} & \num{0} & \num{40} & \num{0} & \num{0}$\pm$\num{0} & \num{0} & \num{10} & \num{0} \\
\cmidrule{3-12}
 &  & \evomasterwbmio{} & \rulePass{} & \num{37.26}$\pm$\num{5.25653598566221} & \num{0} & \num{40} & \num{0} & \num{27.0366666666667}$\pm$\num{5.0495923316689} & \num{0} & \num{0} & \num{10} \\
 &  &  & \ruleFail{} & \num{3.95}$\pm$\num{0.9365309930821} & \num{30} & \num{10} & \num{0} & \num{3.12333333333333}$\pm$\num{0.227737123743285} & \num{0} & \num{0} & \num{10} \\
 &  &  & \ruleWarning{} & \num{6.89666666666667}$\pm$\num{1.12109100389708} & \num{0} & \num{40} & \num{0} & \num{5.38333333333333}$\pm$\num{0.719439081919097} & \num{0} & \num{0} & \num{10} \\
 &  &  & \ruleNotApplied{} & \num{0}$\pm$\num{0} & \num{0} & \num{40} & \num{0} & \num{0}$\pm$\num{0} & \num{0} & \num{10} & \num{0} \\
\cmidrule{3-12}
 &  & \evomasterwbmosa{} & \rulePass{} & \num{37.6}$\pm$\num{3.93594191979335} & \num{0} & \num{40} & \num{0} & \num{28.5033333333333}$\pm$\num{4.6116719873077} & \num{0} & \num{0} & \num{10} \\
 &  &  & \ruleFail{} & \num{4.83}$\pm$\num{0.62908863315212} & \num{20} & \num{20} & \num{0} & \num{3.58666666666667}$\pm$\num{0.369250157768884} & \num{0} & \num{0} & \num{10} \\
 &  &  & \ruleWarning{} & \num{7.1}$\pm$\num{0.300501253482382} & \num{0} & \num{40} & \num{0} & \num{5.40666666666667}$\pm$\num{0.646891904264128} & \num{0} & \num{0} & \num{10} \\
 &  &  & \ruleNotApplied{} & \num{0}$\pm$\num{0} & \num{0} & \num{40} & \num{0} & \num{0}$\pm$\num{0} & \num{0} & \num{10} & \num{0} \\
\cmidrule{3-12}
 &  & \evomasterwbwts{} & \rulePass{} & \num{37.6}$\pm$\num{3.93594191979335} & \num{0} & \num{40} & \num{0} & \num{29.2}$\pm$\num{4.46390185298539} & \num{0} & \num{0} & \num{10} \\
 &  &  & \ruleFail{} & \num{5.23}$\pm$\num{0.545979528448594} & \num{1} & \num{29} & \num{10} & \num{3.65}$\pm$\num{0.207423941139938} & \num{0} & \num{0} & \num{10} \\
 &  &  & \ruleWarning{} & \num{7.1}$\pm$\num{0.300501253482382} & \num{0} & \num{40} & \num{0} & \num{5.51333333333333}$\pm$\num{0.580166004340648} & \num{0} & \num{0} & \num{10} \\
 &  &  & \ruleNotApplied{} & \num{0}$\pm$\num{0} & \num{0} & \num{40} & \num{0} & \num{0}$\pm$\num{0} & \num{0} & \num{10} & \num{0} \\
\bottomrule
\end{tabular}
 \end{table*}
 
This domain-specific \gls{rq} evaluates the tools' ability to execute medical rules.
\Cref{fig:executed-rules} shows the number of (distinct) rules (on the y-axis) by each tool (on the x-axis) for each version of \guri{}.
The dashed line depicts the total number of rules in the particular version.
Each bar represents the arithmetic mean of the rule results \rulePass{}, \ruleFail{}, \ruleWarning{} and \ruleNotApplied{} (see \cref{sec:study:metrics}), with the error bars indicating the standard deviations.
The first two rows depict the validation rules, whereas the last two rows show the aggregation rules.
In addition and similar to the \cref{tab:coverages,tab:errors}, \cref{tab:executed-rules:stats} shows the number of executed rules per tool and rule type aggregated across the versions and repetitions for each result type (\enquote{Mean$\pm$SD}), compares the tool-version pairs with the statistical tests (\enquote{Tool Comp.}, see \cref{sec:study:stats}), and depicts the difference between the \gls{oas} types (\enquote{\gls{oas} Comp.}).

We make four main observations:
\begin{inparaenum}
    \item there is a difference in effectiveness depending on which tool is employed,
    \item the tools are not equally effective for validation and aggregation rules,
    \item the tool effectiveness does not change across the ten versions,
    and
    \item using the \oasStrict{} \gls{oas} achieves equal or better effectiveness for all the tools.
\end{inparaenum}
We discuss these observations in detail below:

\subsubsection{Observation 1: Tool Differences}
\evoguri{} and \evomasterbb{} are on par, executing the most validation rules for \rulePass{} (\num{31.36} and \num{31.36}), \ruleFail{} (\num{37.86} and \num{38.06}), and \ruleWarning{} (\num{0.78} and \num{0.75}).
The statistical tests confirm this, where \evoguri{} and \evomasterbb{} are significantly better than the three \evomasterwb{} tools for \rulePass{} and \ruleFail{} in all the versions and only equal to one another.
For \ruleWarning{}, \evoguri{}, \evomasterbb{}, and \evomasterwbwts{} are equal.
Only for \ruleNotApplied{}, all the tools manage to execute all the rules for it.

The tool choice makes less of a difference for aggregation rules.
All the tools yield \num{37.6} \rulePass{}, about \num{7} \ruleWarning{}{}, and \num{0} \ruleNotApplied{}{} rule results.
We only observe differences for \ruleFail{} results, where \evoguri{}, \evomasterbb{}, and \evomasterwbwts{} outperform \evomasterwbmio{} and \evomasterwbmosa{}.

These results are in line with \rqcodecov{} and \rqfaults{}, i.e., \evoguri{} and \evomasterbb{} are equal or superior to \evomasterwb{}.
Readers accustomed to \gls{rest} \gls{api} test generation research may be surprised, such as \citet{kim:22a}, as white-box tools usually have an advantage over black-box tools.
One reason why the black-box approaches perform better is that the \gls{ea} of \evomasterwb{} optimizes for \enquote{irrelevant} objectives for covering rules.
Once \evomasterwb{} finds better solutions for the source code coverage, it steers itself into a situation where it does not execute more (different) rules anymore.
Conversely, \evoguri{} and \evomasterbb{}, with their simpler approach (concerning covering source code), exercises more diverse inputs (i.e., cancer messages and cases), which leads to more executed rules.

Moreover, we notice that \evoguri{} and \evomasterbb{} have lower variance in the number of executed rules among identical repetitions, whereas all the \evomasterwb{} experience higher variances as indicated by the error bars in \cref{fig:executed-rules}.
This means that the \evomasterwb{} tools cannot consistently execute the same rules for identical repetitions.

\subsubsection{Observation 2: Rule Type Differences}
This leads to the second observation, where the tools are not equally-effective in executing validation rules as they are for aggregation rules.
All the tools execute all the aggregation rules with \rulePass{}, whereas they only manage between \num{11.2} and \num{31.4} of \num{54.1} validation rules.
The reason is inherent to many validation rules, which can only yield a \rulePass{} (as well as \ruleFail{} and \ruleWarning{}) if the left part of an implication is true (see \cref{sec:study:metrics}) and otherwise return a \ruleNotApplied{} result (which all the tools manage for all the rules).
For \ruleFail{} results, no tool manages to execute all the rules for neither validation nor aggregation rules.
However, the tools manage more \ruleFail{} results for validation rules, i.e., \num{18.03} to \num{38.06}, than for aggregation rules, i.e., \num{3.95} to \num{7.1}.
This shows that the tools are more capable of yielding \rulePass{} results for aggregation rules and \ruleFail{} and \ruleNotApplied{} for validation rules;
and, conversely, struggle more to yield \rulePass{} for validation rules and \ruleFail{} and \ruleNotApplied{} for aggregation rules.
They equally struggle to trigger \ruleWarning{} results, which can be explained by it being more challenging finding corner cases or a lack of corner-case-scenarios that lead to \ruleWarning{} results.

We conclude that all the tools follow the same trends in their capabilities and struggles in yielding the four result types.
However, as also outlined before, \evoguri{} and \evomasterbb{} achieve better results than the \evomasterwb{} tools.

\subsubsection{Observation 3: Version Differences}
We observe that, with changing rule sets due to the addition, deletion, and modification of rules, the tools' effectiveness is hardly impacted.
For validation rules, they show increasing number of \rulePass{} and \ruleFail{} with the growing number of rules over the version history;
however, the proportions between the rule results across the versions remains the same.
All the tools manage to \rulePass{} new rules in new versions, but the number of \ruleFail{} and \ruleWarning{} remains constant across the versions.
This suggests that the rule evolution at the \gls{crn} is not an important factor for choosing a particular tool over another.

\subsubsection{Observation 4: \Gls{oas} Differences}
Finally, we compare the number of executed rules and results when using the \oasStrict{} \gls{oas} instead of the \oasDefault{} \gls{oas}.
We observe from \cref{tab:executed-rules:stats} that the \oasStrict{} \gls{oas} considerably raises the number of executed rules with a \rulePass{} and \ruleFail{} result for validation rules for all the tools.
\evoguri{} and \evomasterbb{} also benefit from the \oasStrict{} \gls{oas} for yielding more \ruleWarning{} results.
Aggregation rules are less affected.
\evoguri{} and \evomasterbb{} show an increase in \ruleFail{} results and remain unchanged for the other result types.
The \evomasterwb{} tools benefit more from the \oasStrict{} \gls{oas} for \rulePass{}, \ruleFail{}, and \ruleWarning{} results.
Overall, using the \oasStrict{} \gls{oas} is either equally or more effective than when using the \oasDefault{} \gls{oas}, making it a universal suggestion to employ a \oasStrict{} \gls{oas}.
This is in line with the observations of \citet{kim:22a} that better \gls{oas} lead to more effective test generation.

\summarybox{\rqexecutedrules{} Summary}{
    \evoguri{} and \evomasterbb{} are more effective in generating tests that yield \rulePass{}, \ruleFail{}, and \ruleWarning{} results than the \evomasterwb{} tools.
    While the results are unaffected by the rule evolution, using the \oasStrict{} \gls{oas} leads to considerably better results and can, therefore, be universally suggested for the \gls{crn}.
}

\subsection{\rqrulefrequency{}: Rule and Result Frequency}
\label{sec:results:rule-frequency}

\begin{figure*}[tbp]
    \centering
    \includegraphics[width=\textwidth]{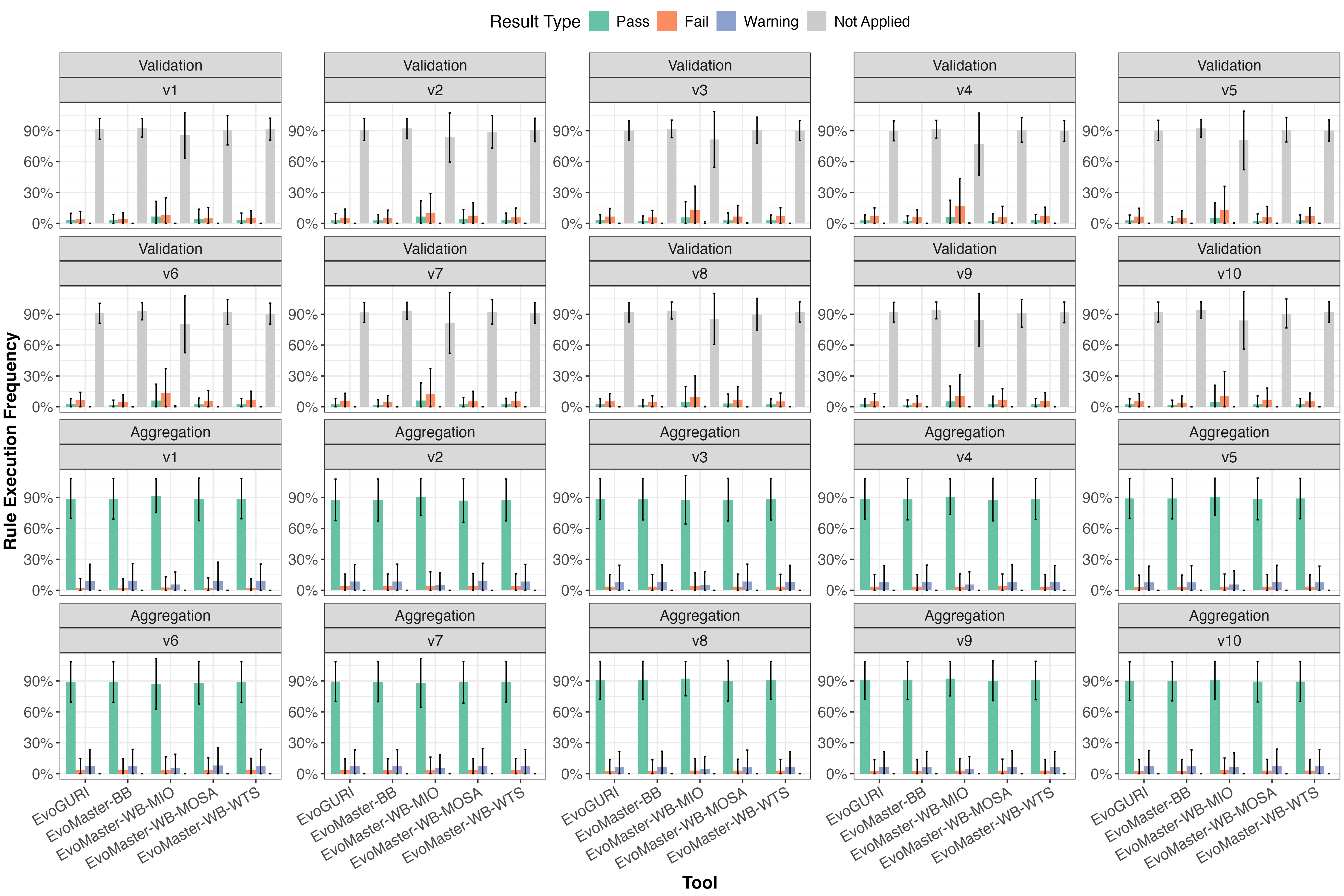}
    \caption{
    Rule execution frequency by result type relative to the total number of rule executions during a test generation phase (i.e., repetition) for each rule, tool, and version.
    The bars are arithmetic means and the error bars are standard deviations.
    }
    \label{fig:rule-execution-results}
\end{figure*}

This domain-specific \gls{rq} evaluates the tools' capabilities to execute rules and their result types frequent during the test generation phase.
The idea is that if a tool executes each rule and result more often, on average across all the rules, it is more likely to generate tests that can expose faults in each rule.
\Cref{fig:rule-execution-results} shows the average rule execution frequency for each result type relative to the total number of rule executions during a test generation phase (i.e., a \qty{1}{\hour} repetition in our experiments) across all the rules (on the y-axis) per tool and result type (on the x-axis), and version (in the facets).
Each bar represents the arithmetic mean, and the error bars are the standard deviations.
The first two rows are for the validation rules, and the last two rows are for the aggregation rules.
In addition, similar to the previous \glspl{rq}, \cref{tab:rule-execution-frequency:stats} depicts the same data aggregated across the versions and provides the results of the statistical tests between the tool-version pairs and the two \gls{oas} types.

We make six observations:
\begin{inparaenum}
    \item \evomasterwbmio{} is most effective in triggering \rulePass{} and \ruleFail{} with \evoguri{} performing competitively;
    \item \ruleNotApplied{} for validation rules and \rulePass{} for aggregation rules are the most frequent results;
    \item there is no effectiveness difference across the versions as in \rqexecutedrules{};
    \item there is a high degree of variance among repetitions;
    \item the \oasStrict{} \gls{oas} improves effectiveness for \rulePass{} and \ruleFail{} results;
    and
    \item the result distribution of the generated tests is different from production \guri{};
\end{inparaenum}

\begin{table*}
    \centering
    \scriptsize
    \caption{
    Rule and result frequencies per tool across all the versions and repetitions, each compared to the other tools in each version and between the \gls{oas} types.
    }
    \label{tab:rule-execution-frequency:stats}
    
    \begin{tabular}{llllrrrlrrr}
\toprule
Rule Type & Tool & Result Type & Mean$\pm$SD & \nrcols{3}{Tool Comp.} & \nrcols{4}{\Gls{oas} Comp.} \\
\cmidrule(l{1pt}r{1pt}){5-7}
\cmidrule(l{1pt}r{1pt}){8-11}
 &  &  &  & $<$ & $=$ & $>$ & Diff. Mean$\pm$SD & $<$ & $=$ & $>$ \\
\midrule
Validation & \evoguriRandom{} & \rulePass{} & \num{2.92483628009967}\%$\pm$\num{0.417134883164084} & \num{1} & \num{33} & \num{6} & \num{1.76539860548012}~\gls{pp}$\pm$\num{0.154251979757414} & \num{0} & \num{0} & \num{10} \\
 &  & \ruleFail{} & \num{5.84054555248775}\%$\pm$\num{0.926041512033138} & \num{1} & \num{33} & \num{6} & \num{3.64314580813181}~\gls{pp}$\pm$\num{1.04771764285386} & \num{0} & \num{0} & \num{10} \\
 &  & \ruleWarning{} & \num{0.00556055331836847}\%$\pm$\num{0.00447690270163115} & \num{0} & \num{28} & \num{12} & \num{0.00556055331836847}~\gls{pp}$\pm$\num{0.00309527737853398} & \num{0} & \num{2} & \num{8} \\
 &  & \ruleNotApplied{} & \num{91.2290576140942}\%$\pm$\num{1.08109813631137} & \num{9} & \num{30} & \num{1} & \num{-5.4141049669303}~\gls{pp}$\pm$\num{1.12745373570309} & \num{10} & \num{0} & \num{0} \\
\cmidrule{2-11}
 & \evomasterbb{} & \rulePass{} & \num{2.39415312735733}\%$\pm$\num{0.403554602299988} & \num{20} & \num{20} & \num{0} & \num{1.2205423033108}~\gls{pp}$\pm$\num{0.138832341474519} & \num{0} & \num{0} & \num{10} \\
 &  & \ruleFail{} & \num{4.78939282789846}\%$\pm$\num{0.781937054010488} & \num{24} & \num{16} & \num{0} & \num{2.56096133366073}~\gls{pp}$\pm$\num{0.801484852008676} & \num{0} & \num{0} & \num{10} \\
 &  & \ruleWarning{} & \num{0.00347311990568405}\%$\pm$\num{0.00297841669353459} & \num{0} & \num{33} & \num{7} & \num{0.00347311990568405}~\gls{pp}$\pm$\num{0.0019306356105194} & \num{0} & \num{2} & \num{8} \\
 &  & \ruleNotApplied{} & \num{92.8129809248385}\%$\pm$\num{0.995034259797803} & \num{0} & \num{13} & \num{27} & \num{-3.78497675687721}~\gls{pp}$\pm$\num{0.875225715979591} & \num{10} & \num{0} & \num{0} \\
\cmidrule{2-11}
 & \evomasterwbmio{} & \rulePass{} & \num{5.75568242354964}\%$\pm$\num{3.80345231415757} & \num{0} & \num{25} & \num{15} & \num{5.1815853358707}~\gls{pp}$\pm$\num{0.541002345625064} & \num{0} & \num{0} & \num{10} \\
 &  & \ruleFail{} & \num{11.6710354919553}\%$\pm$\num{8.82877470415646} & \num{0} & \num{29} & \num{11} & \num{10.3221587553153}~\gls{pp}$\pm$\num{2.8373008601443} & \num{0} & \num{0} & \num{10} \\
 &  & \ruleWarning{} & \num{0.0146240594455194}\%$\pm$\num{0.0912524125312437} & \num{17} & \num{23} & \num{0} & \num{0.0146240594455194}~\gls{pp}$\pm$\num{0.0203451510317518} & \num{0} & \num{10} & \num{0} \\
 &  & \ruleNotApplied{} & \num{82.4077146288232}\%$\pm$\num{12.0632177694581} & \num{13} & \num{27} & \num{0} & \num{14.1686313261255}~\gls{pp}$\pm$\num{8.2430363992903} & \num{0} & \num{10} & \num{0} \\
\cmidrule{2-11}
 & \evomasterwbmosa{} & \rulePass{} & \num{3.01861516830601}\%$\pm$\num{1.82179396747518} & \num{4} & \num{34} & \num{2} & \num{2.9838214463122}~\gls{pp}$\pm$\num{0.68257858570283} & \num{0} & \num{0} & \num{10} \\
 &  & \ruleFail{} & \num{6.20622755814725}\%$\pm$\num{3.41764686534588} & \num{3} & \num{35} & \num{2} & \num{6.14942358370106}~\gls{pp}$\pm$\num{0.621859351542319} & \num{0} & \num{0} & \num{10} \\
 &  & \ruleWarning{} & \num{0.0067377433086364}\%$\pm$\num{0.0303042441482072} & \num{6} & \num{34} & \num{0} & \num{0.0067377433086364}~\gls{pp}$\pm$\num{0.00736000795131965} & \num{0} & \num{10} & \num{0} \\
 &  & \ruleNotApplied{} & \num{90.7684195302381}\%$\pm$\num{4.93796824097907} & \num{3} & \num{33} & \num{4} & \num{4.62598467878372}~\gls{pp}$\pm$\num{5.48272393668661} & \num{0} & \num{10} & \num{0} \\
\cmidrule{2-11}
 & \evomasterwbwts{} & \rulePass{} & \num{2.87797036906161}\%$\pm$\num{0.610555393831509} & \num{2} & \num{34} & \num{4} & \num{2.84764107329718}~\gls{pp}$\pm$\num{0.36173512792057} & \num{0} & \num{0} & \num{10} \\
 &  & \ruleFail{} & \num{6.04994089572082}\%$\pm$\num{1.26208500754725} & \num{0} & \num{31} & \num{9} & \num{5.99052071545952}~\gls{pp}$\pm$\num{0.787443396001059} & \num{0} & \num{0} & \num{10} \\
 &  & \ruleWarning{} & \num{0.00546309864259805}\%$\pm$\num{0.007914509391386} & \num{0} & \num{36} & \num{4} & \num{0.00546309864259805}~\gls{pp}$\pm$\num{0.00299312241098793} & \num{0} & \num{5} & \num{5} \\
 &  & \ruleNotApplied{} & \num{91.066625636575}\%$\pm$\num{1.65150918843554} & \num{8} & \num{31} & \num{1} & \num{-5.07682148259494}~\gls{pp}$\pm$\num{2.9382272824939} & \num{10} & \num{0} & \num{0} \\
\midrule
Aggregation & \evoguriRandom{} & \rulePass{} & \num{89.1891356003393}\%$\pm$\num{0.883801550483855} & \num{0} & \num{31} & \num{9} & \num{0.747530760574366}~\gls{pp}$\pm$\num{0.0951685107362833} & \num{0} & \num{0} & \num{10} \\
 &  & \ruleFail{} & \num{3.17741340155774}\%$\pm$\num{0.40398106713534} & \num{3} & \num{37} & \num{0} & \num{0.552300182816881}~\gls{pp}$\pm$\num{0.06615198969692} & \num{0} & \num{0} & \num{10} \\
 &  & \ruleWarning{} & \num{7.63345099810296}\%$\pm$\num{0.71220985273795} & \num{3} & \num{34} & \num{3} & \num{-1.29983094339125}~\gls{pp}$\pm$\num{0.156806718347753} & \num{10} & \num{0} & \num{0} \\
 &  & \ruleNotApplied{} & \num{0}\%$\pm$\num{0} & \num{0} & \num{40} & \num{0} & \num{0}~\gls{pp}$\pm$\num{0} & \num{0} & \num{10} & \num{0} \\
\cmidrule{2-11}
 & \evomasterbb{} & \rulePass{} & \num{89.0825977763508}\%$\pm$\num{0.896229328702148} & \num{5} & \num{32} & \num{3} & \num{0.660620926355993}~\gls{pp}$\pm$\num{0.0697556958250703} & \num{0} & \num{0} & \num{10} \\
 &  & \ruleFail{} & \num{3.17812222632195}\%$\pm$\num{0.408815760794} & \num{3} & \num{37} & \num{0} & \num{0.56571722187636}~\gls{pp}$\pm$\num{0.065001139995475} & \num{0} & \num{0} & \num{10} \\
 &  & \ruleWarning{} & \num{7.73927999732721}\%$\pm$\num{0.731961786711297} & \num{0} & \num{33} & \num{7} & \num{-1.22633814823235}~\gls{pp}$\pm$\num{0.132858895614362} & \num{10} & \num{0} & \num{0} \\
 &  & \ruleNotApplied{} & \num{0}\%$\pm$\num{0} & \num{0} & \num{40} & \num{0} & \num{0}~\gls{pp}$\pm$\num{0} & \num{0} & \num{10} & \num{0} \\
\cmidrule{2-11}
 & \evomasterwbmio{} & \rulePass{} & \num{90.1948839894886}\%$\pm$\num{8.95951373879413} & \num{0} & \num{23} & \num{17} & \num{65.6794690141401}~\gls{pp}$\pm$\num{9.10993684915618} & \num{0} & \num{0} & \num{10} \\
 &  & \ruleFail{} & \num{3.53170652441704}\%$\pm$\num{0.855409029815909} & \num{0} & \num{34} & \num{6} & \num{2.30536927201112}~\gls{pp}$\pm$\num{0.447572578774728} & \num{0} & \num{1} & \num{9} \\
 &  & \ruleWarning{} & \num{5.36488818283617}\%$\pm$\num{2.22368933480651} & \num{22} & \num{18} & \num{0} & \num{3.53321763667065}~\gls{pp}$\pm$\num{0.925275764531524} & \num{0} & \num{5} & \num{5} \\
 &  & \ruleNotApplied{} & \num{0}\%$\pm$\num{0} & \num{0} & \num{40} & \num{0} & \num{0}~\gls{pp}$\pm$\num{0} & \num{0} & \num{10} & \num{0} \\
\cmidrule{2-11}
 & \evomasterwbmosa{} & \rulePass{} & \num{88.6768909275809}\%$\pm$\num{1.02517927909637} & \num{26} & \num{14} & \num{0} & \num{67.8435859643219}~\gls{pp}$\pm$\num{7.95043582292377} & \num{0} & \num{0} & \num{10} \\
 &  & \ruleFail{} & \num{3.33384154405837}\%$\pm$\num{0.606000260444359} & \num{0} & \num{40} & \num{0} & \num{1.27698102180426}~\gls{pp}$\pm$\num{0.530967522255311} & \num{0} & \num{6} & \num{4} \\
 &  & \ruleWarning{} & \num{7.98926752836073}\%$\pm$\num{1.02687254767735} & \num{0} & \num{25} & \num{15} & \num{5.85222732507846}~\gls{pp}$\pm$\num{1.0164297073929} & \num{0} & \num{4} & \num{6} \\
 &  & \ruleNotApplied{} & \num{0}\%$\pm$\num{0} & \num{0} & \num{40} & \num{0} & \num{0}~\gls{pp}$\pm$\num{0} & \num{0} & \num{10} & \num{0} \\
\cmidrule{2-11}
 & \evomasterwbwts{} & \rulePass{} & \num{89.0710500072447}\%$\pm$\num{0.91819456545143} & \num{3} & \num{32} & \num{5} & \num{69.8773522221596}~\gls{pp}$\pm$\num{4.68761354206651} & \num{0} & \num{0} & \num{10} \\
 &  & \ruleFail{} & \num{3.26800918214465}\%$\pm$\num{0.426557212101528} & \num{0} & \num{40} & \num{0} & \num{0.531539869990911}~\gls{pp}$\pm$\num{0.326193543436954} & \num{0} & \num{10} & \num{0} \\
 &  & \ruleWarning{} & \num{7.66094081061061}\%$\pm$\num{0.768060848383283} & \num{3} & \num{34} & \num{3} & \num{5.75870525829567}~\gls{pp}$\pm$\num{0.633824524277062} & \num{0} & \num{0} & \num{10} \\
 &  & \ruleNotApplied{} & \num{0}\%$\pm$\num{0} & \num{0} & \num{40} & \num{0} & \num{0}~\gls{pp}$\pm$\num{0} & \num{0} & \num{10} & \num{0} \\
\bottomrule
\end{tabular}
 \end{table*}

\subsubsection{Observation 1: Tool Differences}
We observe different results when comparing the tools as for \rqexecutedrules{}.
For validation rules, \evomasterwbmio{}, which was the worst performing tool in \rqexecutedrules{}, achieves the highest frequencies for \rulePass{} (\num{5.76}\%) and \ruleFail{} (\num{11.67}\%), and the lowest for \ruleNotApplied{}.
This is also supported by the statistical tests from \cref{tab:executed-rules:stats}.
This suggests that the most sophisticated search algorithm manages to generate more tests (relatively to all the generated tests) that lead to \rulePass{} and \ruleFail{}, which are arguably the result types of the highest interest to the \gls{crn} developers.
The best performing tools from \rqexecutedrules{}, i.e., \evoguri{} performs mostly equally to and \evomasterbb{} performs often worse than the other tool-version pairs.
For aggregation rules, the tools show mostly similar frequencies to one another with a minor.

\subsubsection{Observation 2: Result Type Differences}
In terms of the four result types, the results show that one result type dominates the execution frequencies.
\ruleNotApplied{} results are most common for validation rules, whereas \rulePass{} results are most common for aggregation rules.
This observation is consistent across the different tools and versions.

\subsubsection{Observation 3: Version Differences}
Similar to \rqexecutedrules{}, we hardly observe any differences across the different versions for all the tools.
This strengthens our suggestion from \rqexecutedrules{} that rule evolution is not a deciding factor for choosing one tool over another.

\subsubsection{Observation 4: Repetition Differences}
The first observation is that there is a high degree of variance of the rule execution frequency across the different rules and repetitions, irrespective of the tool, version, rule type, and result type.
This means that while generating tests, the tools execute the different rules with varying frequency in the different repetitions.
This is caused by how \evoguri{} and \evomaster{} generate (new) variable values of cancer messages and cancer cases: \emph{randomly}.
However, when we compare this with the results from \rqexecutedrules{}, we notice that the frequency variance has hardly any impact on the number of executed rules, i.e., their variance across rules and repetitions is minuscule.

\subsubsection{Observation 5: \gls{oas} Differences}
Similar to \rqexecutedrules{}, employing the \oasStrict{} \gls{oas} largely improves the effectiveness of the tools:
across all the tools, \rulePass{} and \ruleFail{} results are triggered more often than when using the \oasDefault{} \gls{oas}, at the expense of \ruleNotApplied{} results for \evoguri{}, \evomasterbb{}, and \evomasterwbwts{}.
In our previous paper~\citep{laaber:23a}, where we reported the results based on the \oasDefault{} \gls{oas}, \evomasterwbmio{} and \evomasterwbmosa{} often failed to execute rules at all.
This does not happen with the \oasStrict{} \gls{oas} anymore.
Overall, we can conclude that relying on the \oasStrict{} \gls{oas} positively affects the rule execution frequency.

\subsubsection{Observation 6: Comparison to Production \guri{}}

\begin{table*}
    \centering
    \scriptsize
    \caption{
    Rule execution frequency by result type relative to the total number of rule executions during a test generation phase (i.e., repetition) for each rule and tool.
    The values are arithmetic means $\pm$ standard deviations summarized on rule type level across all the individual rules.
    \tool{Production} corresponds to the rule executions from production \guri{}.
    The values for the test generation tools are for the latest version \version{v10}, averaged across all the repetitions.
    }
    \label{tab:rule-execution-frequency:production}
    
    \begin{tabular}{llllll}
\toprule
Rule Type & Tool & \nrcols{4}{Result Type} \\
\cmidrule(l{1pt}r{1pt}){3-6}
 &  & Pass & Fail & Warning & Not Applied \\
\midrule
Validation & \tool{Production} & \num{26.1918255485415}\%$\pm$\num{37.3160629293796} & \num{0.0535950998398514}\%$\pm$\num{0.153859882156909} & \num{6.55323984204683e-06}\%$\pm$\num{5.52185804086332e-05} & \num{73.7545727983788}\%$\pm$\num{37.3785677226814} \\
 & \evoguriRandom{} & \num{2.58556587178214}\%$\pm$\num{5.19723214874637} & \num{5.26963727115051}\%$\pm$\num{7.61941456871123} & \num{0.0056964003012674}\%$\pm$\num{0.0530840989514006} & \num{92.1391004567661}\%$\pm$\num{9.62587361566069} \\
 & \evomasterbb{} & \num{2.0271726198982}\%$\pm$\num{4.45884683345244} & \num{4.1140349139415}\%$\pm$\num{6.40559761094812} & \num{0.00355610297677561}\%$\pm$\num{0.0347973835746043} & \num{93.8552363631835}\%$\pm$\num{8.05753805784333} \\
 & \evomasterwbmio{} & \num{5.06963833033936}\%$\pm$\num{15.9325793023213} & \num{10.8707676218355}\%$\pm$\num{23.6178382084016} & \num{0.00364455992172075}\%$\pm$\num{0.122208319276589} & \num{84.0559494879034}\%$\pm$\num{27.9143178935254} \\
 & \evomasterwbmosa{} & \num{2.82878383760782}\%$\pm$\num{7.76788805883441} & \num{6.45809413105794}\%$\pm$\num{11.7359367732925} & \num{0.0059345783966608}\%$\pm$\num{0.127686834430368} & \num{90.7071874529376}\%$\pm$\num{14.0256138752877} \\
 & \evomasterwbwts{} & \num{2.5791548117516}\%$\pm$\num{5.335332822714} & \num{5.31544489755883}\%$\pm$\num{8.06167950105593} & \num{0.00669280688250784}\%$\pm$\num{0.0797846756492714} & \num{92.098707483807}\%$\pm$\num{9.94461111501391} \\
\midrule
Aggregation & \tool{Production} & \num{99.8835781985487}\%$\pm$\num{0.472240433595079} & \num{0.0188443056469122}\%$\pm$\num{0.0818327448630621} & \num{0.097577495804357}\%$\pm$\num{0.465327357318205} & \num{0}\%$\pm$\num{0} \\
 & \evoguriRandom{} & \num{89.8076725462932}\%$\pm$\num{18.7197816516764} & \num{2.87933980667846}\%$\pm$\num{10.614452146497} & \num{7.31298764702837}\%$\pm$\num{15.4147672474053} & \num{0}\%$\pm$\num{0} \\
 & \evomasterbb{} & \num{89.6701368911985}\%$\pm$\num{18.9669065880754} & \num{2.90321456508029}\%$\pm$\num{10.6337234747816} & \num{7.42664854372118}\%$\pm$\num{15.6597291614518} & \num{0}\%$\pm$\num{0} \\
 & \evomasterwbmio{} & \num{90.5722512266354}\%$\pm$\num{18.5376557343589} & \num{3.30978290944074}\%$\pm$\num{11.8645476086069} & \num{6.11796586392385}\%$\pm$\num{14.1272661969257} & \num{0}\%$\pm$\num{0} \\
 & \evomasterwbmosa{} & \num{89.3763456432167}\%$\pm$\num{19.6992220466371} & \num{3.01665201370771}\%$\pm$\num{10.9425985388986} & \num{7.60700234307564}\%$\pm$\num{16.262308333883} & \num{0}\%$\pm$\num{0} \\
 & \evomasterwbwts{} & \num{89.4903187855807}\%$\pm$\num{19.3286079052846} & \num{2.99162027569604}\%$\pm$\num{10.7908234113267} & \num{7.51806093872324}\%$\pm$\num{15.8902008521014} & \num{0}\%$\pm$\num{0} \\
\bottomrule
\end{tabular}
 \end{table*}
 
Finally, we compare the tools' rule execution and result frequency to those from production \guri{}, i.e., to real-world statistics of rule execution results.
\Cref{tab:rule-execution-frequency:production} depicts this for \enquote{Production} and compares it to all the five tools.
The values are arithmetic means and standard deviations relative to the total number of rules.
Note that we only consider the tool results for the latest version, i.e., \version{v10}, as the production results are only available for the current version.

We observe that the distributions of the tools are considerably different from production \guri{}.
In production, most rules are \ruleNotApplied{} (\num{73.75}\%) followed by \rulePass{} results (\num{26.19}\%).
Only a fraction fails, and even fewer yield a warning.
The situation is even more extreme for aggregation rules: \num{99.88}\% of the rules \rulePass{}, and only a negligible number \ruleFail{} or yield a \ruleWarning{}.
The closest tool to production, in terms of effectiveness, is \evomasterwbmio{}.
However, none of the tools achieves a similar number of rules that \rulePass{} or \ruleFail{} for both rule types.

Interestingly, the tools often achieve higher \ruleFail{} and \ruleWarning{} rates than in production.
In particular, \ruleFail{} results occur considerably more often with \num{4.11}\% to \num{10.87}\% for validation rules and \num{2.88} to \num{3.31}\% for aggregation rules.
The tools trigger \ruleWarning{} results in around \num{0.01}\% of the validation rule executions and between \num{6.12}\% and \num{7.61}\% for aggregation rules.
This shows that all the tools are potent in many generating tests for failure and corner cases that lead to \ruleFail{} and \ruleWarning{} results.

\summarybox{\rqrulefrequency{} Summary}{
    \evomasterwbmio{} has the highest number of \rulePass{} and \ruleFail{} results with \evoguri{} being competitively.
    For all the tools, the validation rules are the easiest to yielding \ruleNotApplied{} followed by \ruleFail{}, and aggregation rules are the easiest to return \rulePass{} followed by \ruleWarning{}.
    The \oasStrict{} \gls{oas} improves effectiveness across all the tools, especially for \rulePass{} and \ruleFail{} results.
    Compared to production \guri{}, no tool can \rulePass{} a similar amount of rules, but all are good at generating tests that lead to \ruleFail{} and \ruleWarning{} results;
}

\section{\rqmut{}: Mutation Testing}
\label{sec:results:mut}

\begin{table}
    \centering
    \scriptsize
    \caption{
    Mutation testing results.
    }
    \label{tab:mutation-testing}
    
    \begin{tabular}{llrrr}
\toprule
Tool & \Gls{oas} Type & Tests & $MS^V$ & $MS^A$ \\
\midrule
\evoguriRandom{} & \oasStrict{} & \num{43} & \num{0.956896551724138} & \num{1} \\
 & \oasDefault{} & \num{19} & \num{0} & \num{1} \\
\evomasterbb{} & \oasStrict{} & \num{10} & \num{0} & \num{0} \\
 & \oasDefault{} & \num{8} & \num{0} & \num{0} \\
\evomasterwbmio{} & \oasStrict{} & \num{34} & \num{0.125} & \num{1} \\
 & \oasDefault{} & \num{13} & \num{0} & \num{1} \\
\evomasterwbmosa{} & \oasStrict{} & \num{48} & \num{0.964080459770115} & \num{1} \\
 & \oasDefault{} & \num{17} & \num{0} & \num{1} \\
\evomasterwbwts{} & \oasStrict{} & \num{50} & \num{0.956896551724138} & \num{1} \\
 & \oasDefault{} & \num{20} & \num{0} & \num{1} \\
\bottomrule
\end{tabular}
 \end{table}
 
Finally, this \gls{rq} evaluates the ability in finding artificial rule faults by the tools' generated test suites.
Based on the \numMutationOperators{} rule mutation operators, the experiment comprises \numMutations{} rule mutations against which the test suites are executed (see \cref{sec:study:mutations,sec:study:setup}).
\Cref{tab:mutation-testing} depicts the \glspl{ms} per tool and \gls{oas} type, i.e., separating the \gls{ms} for validation rules $MS^V$ and aggregation rules $MS^A$.

We observe that \evoguri{}, \evomasterwbmosa{}, and \evomasterwbwts{} achieve the highest \gls{ms} for validation rules, i.e., \num{0.96}, almost killing all the mutants.
For aggregation rules, \evoguri{} and the \evomasterwb{} tools achieve a perfect score of \num{1}, killing all the mutants.
A reason for the high \glspl{ms} is that the mutations are relatively easy to kill, as a minor change to the rule already yields a different result and can be detected by the test suite.

Interestingly, \evomasterbb{}, which is tied for the number of executed rules with \evoguri{} in \rqexecutedrules{}, has a \gls{ms} of \num{0} for both validation and aggregation rules.
We assume that this happens because of the low number of retained test cases in the final test suite.
All the other tools generate larger test suites, as they aim for covering more search targets than \evomasterbb{} (see \cref{sec:study:tools}).
Actually, the increased number of search targets is the key difference between \evomasterbb{} and \evoguri{}.

Similar to the other \glspl{rq}, employing the \oasStrict{} \gls{oas} enables killing mutants at all for validation rules.
This is in line with our previous paper's results~\citep{laaber:23a} and the \oasDefault{} \gls{oas} results from \rqexecutedrules{}, where hardly any rules yielded \rulePass{}, \ruleFail{}, or \ruleWarning{} and most are \ruleNotApplied{}; consequently, the generated tests are incapable of killing the mutants.
For aggregation rules, the \gls{oas} does not impact the \glspl{ms}, as all the tools are able to \rulePass{} all the rules.

\summarybox{\rqmut{} Summary}{
    \evoguri{}, \evomasterwbmosa{}, and \evomasterwbwts{} are highly effective in killing mutants for both validation and aggregation rules, with \evomasterbb{} failing to detect the mutants.
    The \oasStrict{} \gls{oas} has a big impact on the \gls{ms} for validation rules.
    We can suggest \evoguri{} being used at the \gls{crn} due to its high \glspl{ms} and simpler setup than the \evomasterwb{} tools.
}
 
\section{Discussion and Lessons Learned}
\label{sec:discussion}

This section discusses the results and outlines lessons learned, which provide research opportunities.

\subsection{Need for Domain-Specific Objectives, Targets, and Evaluation Metrics}

From \rqcodecov{} and \rqfaults{}, we see that the differences among the tools are minor, even though they may be statistically significant.
Hence, we require domain-specific evaluation metrics to differentiate between the tools when traditional evaluation metrics fail to do so.
Therefore, we introduced two novel domain-specific metrics in \rqexecutedrules{} and \rqrulefrequency{} and adapted a well-known metric, i.e., \gls{ms}, to our domain-specific case in \rqmut{}.
Moreover, influenced by the findings of our previous paper~\citep{laaber:23a}, we extended the best tool, i.e., \evomasterbb{}, and incorporate domain-specific targets (rule and result types) into the test generation process with \evoguri{}.
Our paper shows that domain-specificity is essential to evaluating current tools and adapting them for real-world contexts.
Hence, going forward, test generation tools require
\begin{inparaenum}
    \item encoding domain-specific objectives in their search, e.g., with added domain-specific search objectives (e.g., rule (result) count or distance metrics to applying rules~\citep{ali:13}), or adding tests that reach unseen domain-specific targets to the archive (similar to \citet{padhye:19a});
    \item keeping tests after the search that cover each domain-specific target for regression testing scenarios;
    and
    \item evaluating test generation tools with domain-specific metrics (rule execution status and rule execution results in our case) to show their effectiveness when traditional metrics do not show differences (also discussed by \citet{boehme:22}).
\end{inparaenum}

\subsection{Oracle Problem for Rule Execution Results}

The oracle problem is a well-known problem in software testing~\citep{barr:15b}, which extends also to domain-specific goals.
While we show that the current tools can execute rules with different results, it is unclear if, for a randomly generated test input, a rule is expected to \rulePass{}, \ruleFail{}, yield a \ruleWarning{}, or should be \ruleNotApplied{}.
Current research does not offer a solution; there simply is no implicit oracle~\citep{barr:15b} for rule executions.
Going forward, this needs to be addressed, and we see four potential aspects:
\begin{inparaenum}
    \item using tests that lead to a specific rule result (\rulePass{}, \ruleFail{}, \ruleWarning{}) and employing them in a regression testing setting, e.g., if a test passes a rule in version \version{1}, it should also pass in version \version{2}, for which one could use the test suites generated by \evoguri{};
    \item applying differential testing by comparing the outputs of the same random test input to, e.g., a reference implementation for the medical rules, as they should be standardized, which we explored in \citet{isaku:24};
    \item devising metamorphic relations on the rules that are either semantics-preserving (similar to \citet{lu:19}) or are known to lead to invalid rules and comparing the outputs to the correct implementation;
    and
    \item using \glspl{llm} as oracles, as they are likely trained on the medical knowledge, regulations, and standards, by prompting whether a rule result returned by \guri{} is correct for a given input.
\end{inparaenum}

\subsection{Challenge of Generating Medical Data}

Generating synthetic medical data is a challenge for researchers from many fields are trying to tackle~\citep{Goncalves2020, Dankar2021}.
In software testing, generating synthetic and valid medical data is equally important.
The current test generation tools are good at generating syntax-compliant data (e.g., according to an OpenAPI schema definition).
However, the individual variable values (i.e., medical variables) are randomly generated.
This leads to many invalid cancer messages and cases to be checked by the rule engine, such as when using the \oasDefault{} \gls{oas} as done in our previous paper~\citep{laaber:23a}.
We improved on this in this paper by adding a \oasStrict{} \gls{oas} that constrains the medical variables to only valid values, which we manually derived from \gls{crn}-internal documentation, which improves effectiveness across all our \glspl{rq}.
However, this only leads to valid data and not necessarily to realistic data, i.e., in our context, data that accurately mimics real cancer patient data.
We identify four potential ways to address this challenge in the context of test generation in the future:
\begin{inparaenum}
    \item using generative models such as \glspl{gan} or \glspl{vae} trained on real patient records to generate valid and realistic cancer messages and cases to directly test \guri{}~\citep{Hernandez2022};
    \item using generated cancer messages and cases in \evomaster{} either as seeds or when new requests are sampled;
    and
    \item employing \glspl{llm}, possibly trained on \glspl{ehr} or medical text~\citep{singhal:22,lievin:23,yang:22,yunxiang:23}, to generate variable values.
\end{inparaenum}

\subsection{Generality of the Results}

While the results are specific to the case study, i.e., \guri{} at the \gls{crn}, the findings and lessons learned are likely applicable in other contexts.
\begin{inparaenum}
    \item Other countries also have medical registries similar to the \gls{crn}, which also deals with \gls{ehr}.
    These can benefit from the challenges and findings outlined in this paper to introduce automated test generation tools.
    \item Rule-based systems, in general, potentially face similar challenges.
    Once generated tests pass the input validation and execute rules, they will also have to deal with, e.g., domain-specific objectives, the oracle problem, and generating data that executes the rules.
    \item Any system that dynamically loads targets of interest will likely also be affected by the need for domain-specific objectives and targets.
    \item Researchers at a recent Dagstuhl seminar discussed similar challenges, such as domain-specific objectives and targets; comparison to production; domain-specific oracles such as reference implementations, differential testing, and metamorphic testing; and required evaluations that go beyond code coverage and errors~\citep{boehme:23}.
\end{inparaenum}
This shows that while our results are specific to \guri{}, the challenges are important to the research community.
Consequently, our paper is a valuable case study providing data for these challenges.

\subsection{Call for Studies with Domain-Specific Goals}

Based on our findings showing that code coverage and uncovered errors are insufficient to evaluate automated test generation tools, we conclude that there is a dire need for more industrial and public sector case studies like ours.
As also outlined by \citet{boehme:22}, code coverage is insufficient to validate test generation tools and fuzzers.
For example, which tool is better when code coverage is equal, or no errors are found?
Exactly this happens in the \gls{crn}'s case.
The research community needs to better understand domain-specific needs for test generation, objectives, and targets to optimize for, and evaluation metrics that are better aligned with stakeholders' interests; and, as a next step, evolve current tools to support these domain-specific needs better.
 
\section{Related Work}
\label{sec:rw}

\subsection{Test Generation for \gls{rest} \glspl{api}}

Many industrial applications, especially those built with the microservice architecture expose \gls{rest} \glspl{api}.
As a result, there is an increasing demand for automated testing of such \gls{rest} \glspl{api}.
Consequently, we can see a significant rise in publications in recent years~\citep{golmohammadi:23}.
Moreover, several open-source and industrial \gls{rest} \gls{api} testing tools are available such as
\evomaster{}~\citep{arcuri:19,arcuri:18a,arcuri:18b,arcuri:21}, 
RESTler~\citep{atlidakis:19}, 
RestTestGen~\citep{corradini:22},
RESTest~\citep{martin-lopez:21b}, 
Schemathesis~\citep{schemathesis},
Dredd~\citep{dredd},
Tcases~\citep{tcases},
bBOXRT~\citep{laranjeiro:21,evorefuzz},
and
APIFuzzer~\citep{apifuzzer}.
Even though any of these tools can be used in our context, we use \evomaster{}, since it is open-source and has been shown to be the most effective regarding source code coverage and thrown errors among ten different tools in a recent study~\citep{kim:22a}.

Generally, \gls{rest} \gls{api} testing approaches are classified into black-box (no source code access) and white-box (requires source code access)~\citep{golmohammadi:23}.
The existing literature has developed testing techniques from three main perspectives to evaluate testing effectiveness~\citep{golmohammadi:23}:
\begin{inparaenum}
    \item coverage criteria, e.g., code coverage (e.g., branch coverage) and schema coverage (e.g., request input parameters);
    \item fault detection, e.g., service errors (i.e, \gls{http} status code 5XX) and \gls{rest} \gls{api} schema violations;
    and
    \item performance metrics, i.e., related to the response time of \gls{rest} \gls{api} requests.
\end{inparaenum}
To achieve these objectives, various algorithms have been developed in the literature.
For instance, \evomaster{} has implemented several \glspl{ea} including random testing~\citep{arcuri:19,arcuri:18b,arcuri:21}.
Various extensions have been also proposed to \evomaster, such as handling sequences of \gls{rest} \gls{api} calls and their dependencies~\citep{zhang:21b}, handling database access through \gls{sql}~\citep{zhang:21a}, and testing \gls{rpc}-based \glspl{api}~\citep{zhang:23b}.
In our case, we have access to the source code of \guri{}; therefore, we employ both the black-box and the white-box (parameterized with three \glspl{ea}) tools of \evomaster{}.
However, an extended investigation in the future may include other tools.

Compared to the literature, our main contribution is applying an open-source \gls{rest} \gls{api} testing tool in the real-world context of the \gls{crn}.
We assess the tool's effectiveness in achieving code coverage, errors found, and domain-specific metrics (e.g., related to medical rules defined for validating and aggregating cancer messages and cancer cases).

\subsection{Development and Testing of Cancer Registry Systems}
In our recent paper with the \gls{crn}~\citep{laaber:23b}, we assess the current state of practice and identify challenges (e.g., test automation, testing evolution, and testing \gls{ml} algorithms) when testing \gls{crn}['s] \gls{caress}.
In our previous paper~\citep{laaber:23a}, we took the first concrete step towards handling those challenges, particularly assessing the effectiveness of an existing testing tool in the \gls{crn}'s context.
In this paper, we extend our previous paper based on our findings and introduce a domain-specific extension to \evomaster{} named \evoguri{}, refine the \gls{oas} to only generate valid cancer messages and cases, and assess the fault-finding capabilities of the generated test suites with a domain-specific mutation testing approach.
Three other recent works build cyber-cyber physical twins for \guri{}~\citep{lu:23}, incorporate \gls{ml} classifiers into \evomaster{} to reduce testing cost~\citep{isaku:23}, and utilize \glspl{llm} to generate test cases used in a differential testing approach with a reference implementation~\citep{isaku:24}.

In the past, we developed a model-based engineering framework to support \gls{caress} at the \gls{crn}~\citep{wang:16b}.
The framework aims to create high-level and abstract models to capture various rules, their validation, selection, and aggregation.
The framework is implemented based on the \gls{uml} and \gls{ocl}, where the \gls{uml} is used to capture domain concepts and the \gls{ocl} is used to specify medical rules. 
The implementation of the framework has been incorporated inside \guri{}, which is the subject of testing in this paper.
As a follow-up, we also developed an impact analysis approach focusing on capturing changes in rules and assessing their impact to facilitate a systematic evolution of rules~\citep{wang:17}.
Finally, we also developed a search-based approach to refactor such rules regarding their understandability and maintainability~\citep{lu:19}.
Compared to the existing works, this paper focuses on testing \guri{} as a first step toward building cost-effective testing techniques at the \gls{crn}.

\section{Conclusions}
\label{sec:conclusions}
\glsresetall{}

This paper reports on an empirical study evaluating the test effectiveness of \evomaster{}'s four testing tools and \evoguri{} a domain-specific \evomaster{} extension on a real-world system, the \gls{crn}'s \gls{caress}.
\Gls{caress} is a complex software system that collects and processes cancer patients' data in Norway and produces statistics and data for its end users.
Our results show that the five studied testing tools perform similarly regarding code coverage and errors reported across all the studied versions.
However, in terms of domain-specific metrics, our \gls{crn}-specific \evoguri{} is the most effective tool; hence, we recommend it for the \gls{crn} as an automated test generation solution.
We also provide lessons learned that are beneficial for researchers and practitioners.

\section*{Acknowledgements}
This work received funding from \gls{rcn} under the project \texttt{309642} and has benefited from the \glsentryfull{ex3}, which is supported by the \gls{rcn} project \texttt{270053}.

\footnotesize
\bibliographystyle{IEEEtranSN}
\bibliography{refs.bib,refs-jan.bib}

\end{document}